**Title:** Evaluation and comparison of MODIS Collection 6.1 aerosol optical depth against AERONET over regions in China with multifarious underlying surfaces

**Authors:** Yuan Wang [a], Qiangqiang Yuan [a,e,f,*], Tongwen Li [b], Huanfeng Shen [b,d,f], Li Zheng [a], Liangpei Zhang [c,f]

**Affiliations:**

[a] School of Geodesy and Geomatics, Wuhan University, Wuhan, Hubei, 430079, China.

[b] School of Resource and Environmental Sciences, Wuhan University, Wuhan, Hubei, 430079, China.

[c] The State Key Laboratory of Information Engineering in Surveying, Mapping and Remote Sensing, Wuhan University, Wuhan, Hubei, 430079, China.

[d] The Key Laboratory of Geographic Information System, Ministry of Education, Wuhan University, Wuhan, Hubei, 430079, China.

[e] The Key Laboratory of Geospace Environment and Geodesy, Ministry of Education, Wuhan University, Wuhan, Hubei, 430079, China.

[f] The Collaborative Innovation Center for Geospatial Technology, Wuhan, Hubei, 430079, China.

**Email addresses:**

whuwy@yahoo.com (Wang Yuan), yqiang86@gmail.com (Qiangqiang Yuan), litw@whu.edu.cn (Tongwen Li), shenhf@whu.edu.cn (Huanfeng Shen), lzheng@sgg.whu.edu.cn (Li Zheng), zlp62@whu.edu.cn (Liangpei Zhang)

[*] **Corresponding author:**

Qiangqiang Yuan (yqiang86@gmail.com). Phone: +86-027-68758427

**ABSTRACT**

In this study, we evaluated the performance of the Moderate Resolution Imaging Spectroradiometer (MODIS) Collection 6.1 (C6.1) aerosol optical depth (AOD) products and compared them with the Collection 6 (C6) products over regions in China with multifarious underlying surfaces during 2001–2016. The AOD retrievals were validated against 20 AERONET (V3) sites, and the results show that the correlation coefficient (R) for dark target (DT) retrievals in C6.1 is 0.946, while the fraction within the expected error (EE) can be considered relatively low at only 54.03%. Deep blue (DB) retrievals in C6.1 have a slightly lower R value (0.931), but the other criteria are superior to DT. Comparing the results over urban and vegetation areas in C6.1, the overall quality of the DB retrievals is better than the DT retrievals in urban areas. The performance of DT is significantly superior to DB in the low elevation vegetation (LEV) areas. For the high elevation vegetation (HEV) areas, DB performs synthetically better than DT. In the spatial distribution aspect in C6.1, most of the DB AOD values are less than those of DT, and the relationship between DT and DB varies with the different land cover types. For the AOD coverage in C6.1, DT retrievals with high coverage mainly distribute in east-central China. However, the effects of high surface reflectance lead to low AOD coverage in the southwest. In contrast, the AOD coverage of DB tends to be high in areas where the main land cover type is bare soil and tends to be low in areas affected by snow. In terms of the comparison between C6.1 and C6, the overestimation of DT over urban areas in C6 is effectively mitigated in C6.1. However, a nearly systematic decline in DT is discovered in C6.1 as well. With respect to DB, consistent AOD coverage distribution is observed, with only subtle distinction. The AOD coverage of DB in C6.1 appears higher than that in C6 in the middle, south, and northeast of China. The quality of the DB retrievals in C6.1 increases slightly compared to C6, and the most remarkable improvement is observed for the coarse aerosol particles.

# 1. Introduction

As a major source of uncertainty in the atmosphere, aerosol in the global climate system varies both spatially and temporally (IPCC, 2014; Kaufman et al., 2002). By interacting with solar and terrestrial radiation and altering cloud properties and lifetime, aerosol can lead to radiative forcing (Bellouin et al., 2005). Aerosol not only has an effect on atmospheric radiation, but also air quality (Pope III et al., 2009). Particulate matter with an aerodynamic equivalent diameter (AED) of less than 2.5 μm ($PM_{2.5}$) can carry toxic and noxious substances and transport them across countries and geographic boundaries (Brauer et al., 2015; Yang et al., 2017). The impact of aerosol that stems from natural and anthropogenic sources in air pollution has drawn significant attention from researchers (Volkamer et al., 2006). Typically, a key parameter called the "aerosol optical depth" (AOD) is utilized to depict aerosol optical properties (Holben et al., 2001; Xin et al., 2007). On account of the drawback of ground-based measurements, which fail to offer a global perspective (Kaufman et al., 2005), various satellites have been applied to retrieve AOD (Kahn et al., 2010). As the most widely used products, AOD retrievals from the Moderate Resolution Imaging Spectroradiometer (MODIS) on board the Terra and Aqua satellites are favored among scholars. With 36 spectral channels, a satisfactory temporal resolution of 1 to 2 days, and moderate spatial resolutions of 250 m, 500 m, and 1000 m (Bisht et al., 2005; Justice et al., 1998; King et al., 1992), MODIS can generate regular AOD products which are conducive to the understanding of the effects of atmospheric aerosols at both local and global scales.

Despite the fact that the Terra and Aqua satellites have operated for more than a decade (Barnes et al., 2003), the enthusiasm of NASA to upgrade the algorithms is still undiminished. From Collection 5.1 (C5.1) released in 2008 to Collection 6 (C6) released in 2012, the MODIS Adaptive Processing System (MODAPS) provides global AOD products over land at 3-km or 10-km spatial resolutions based on the dark target (DT) algorithm and a 10-km spatial resolution based on the deep blue (DB) algorithm (Bilal et al., 2016; He et al., 2010; Kumar et al., 2018; Mhawish et al., 2017; Sayer et al., 2014; Tao et al., 2015; Wang et al., 2010; Wei et al., 2017; Xie et al., 2011). MODAPS is currently generating an improved Collection 6.1 (C6.1) for all MODIS L1 and higher-level atmospheric products and plans to generate C6.1 for all the land products in the near future. Compared to C6, the DT algorithm which modifies the AOD products over land surfaces when the urban percentage (UP) is greater than 20% uses a revised surface characterization (Levy et al., 2016) in C6.1. With regard to the DB algorithm, several principal improvements, i.e., artifact reduction in heterogeneous terrain and improved surface modeling in elevated terrain, are introduced in C6.1. Moreover, the heavy smoke detection and assumed aerosol optical models of region or season have also been updated in the DB algorithm (Hsu, 2017). As C6.1 has only recently been released, detailed research on the new AOD products is scarce and more in-depth analysis is required.

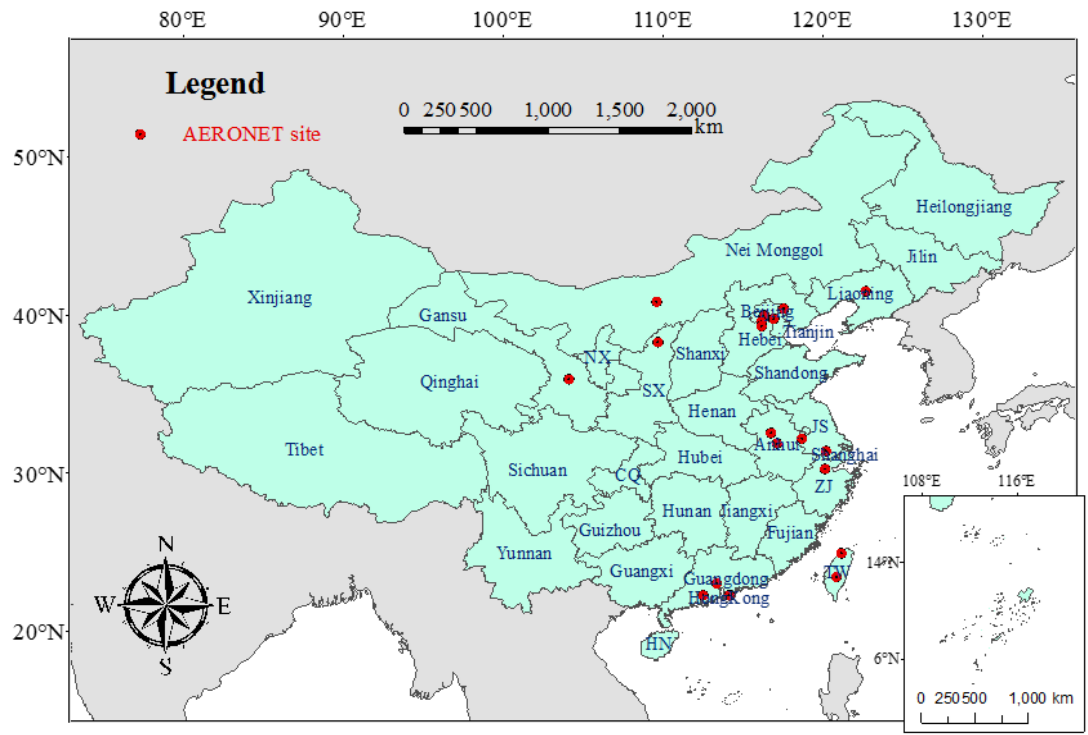

**Fig. 1** Distribution of the AERONET sites considered in this study. The detailed information of each site can be found in Table 1. The name of each province in China is used for analysis in Section 3. NX: Ningxia; SX: Shaanxi; CQ: Chongqing; HN: Hainan; JS: Jiangsu; ZJ: Zhejiang; TW: Taiwan.

China is the largest developing country in the world. Nevertheless, since the urbanization and industrialization of China have accelerated during the past few decades, environmental problems such as air pollution are increasingly emerging. In China, the use of MODIS AOD products is widespread in environmental protection, where the primary application is to estimate the air quality by mapping $PM_{2.5}$ from local to nationwide scales (Li et al., 2017a, 2017b). The error of the MODIS AOD products compared to the AEROsol Robotic NETwork (AERONET) can cause deviation in the estimation of $PM_{2.5}$ (Van Donkelaar et al., 2010). Therefore, it is important to ensure the quality and evaluate the performance of the C6.1 AOD products for China. Furthermore, the properties of aerosol are known to be correlated to surface types (Wei et al., 2017; Zou et al., 2016) and geographical conditions (He et al., 2016; Li et al., 2003; Wang et al., 2010) in China. In the past, many scholars have undertaken excellent research to evaluate the AOD products for China (He et al., 2010, 2017; Tao et al., 2015; Wang et al., 2010; Wei et al., 2017; Xie et al., 2011), yet no one has simultaneously considered the specific land cover types and the digital elevation model (DEM). In conclusion, it is necessary and meaningful to utilize land cover type products (e.g., MCD12Q1) and DEM products (e.g., SRTM1 V3.0) when evaluating the AOD products for China. Meanwhile, the AERONET collaboration updated the processing algorithms from Version 2.0 to Version 3.0 (Giles et al., 2017). Hence, in this study, we employed the Version 3 AERONET databases to enhance the reliability of the validation.

The purpose of this study was to evaluate the MODIS AOD products in C6.1 and compare them with the C6 products over regions in China with multifarious underlying surfaces during 2001–2016. Due to the similar performance of the AOD products from Terra and Aqua (Sayer et al., 2014; Wang et al., 2010), only the Terra AOD product with a 10-km spatial resolution (MOD04L2) was adopted. The rest of this paper is structured as follows. Section 2 introduces the datasets and methodology. The evaluation results and discussion are provided in section 3. Finally, Section 4 provides the conclusion.

## 2. Datasets and methodology

*2.1. Datasets*

In our study, DT and DB retrievals with a 10-km spatial resolution in C6.1 and C6 from Terra during 2001–2016 were utilized. In order to acquire a well-rounded analysis, a land cover type product (MCD12Q1) and a DEM product (SRTM1 V3.0) were also employed. The AOD retrievals were validated using Version 3.0 Level 2.0 AERONET AOD products from ground sites in China.

*2.1.1 AEROsol Robotic NETwork (AERONET)*

**Table 1:** Detailed information about the AERONET sites considered in this study. Lat: latitude; Lon: longitude; Ele: elevation.

| Main surface type | Site | Lat/Lon | Ele (m) | Period |
|---|---|---|---|---|
| Urban | Beijing (BJ) | 39.977/116.381 | 92 | 2001–2016 |
|  | Beijing-CAMS (BJC) | 39.933/116.317 | 106 | 2012–2016 |
|  | Zhongshan_Univ (ZSU) | 23.060/113.390 | 27 | 2011–2012 |
| High elevation vegetation | Lulin (LL) | 23.469/120.874 | 2868 | 2006–2016 |
|  | SACOL (SA) | 35.946/104.137 | 1965 | 2006–2013 |
|  | Yulin (YL) | 38.283/109.717 | 1080 | 2001–2002 |
|  | AOE_Baotou (ABT) | 40.852/109.629 | 1270 | 2013 |
|  | Xinglong (XL) | 40.396/117.578 | 970 | 2006–2014 |
| Low elevation vegetation | Shouxian (SX) | 32.558/116.782 | 22 | 2008 |
|  | Hangzhou_City (HZC) | 30.290/120.157 | 30 | 2008–2009 |
|  | NUIST (NU) | 32.206/118.717 | 62 | 2008–2010 |
|  | Liangning (LN) | 41.512/122.701 | 15 | 2005 |
|  | Hefei (HF) | 31.905/117.162 | 36 | 2005–2008 |
|  | PKU_PEK (PP) | 39.593/116.184 | 66 | 2006, 2008 |
|  | XiangHe (XH) | 39.754/116.962 | 36 | 2004–2016 |
|  | Yufa_PEK (YP) | 39.309/116.184 | 20 | 2006 |
|  | Kaiping (KP) | 22.315/112.539 | 51 | 2008 |
| Water | Hong_Kong_PolyU (HKP) | 22.303/114.180 | 30 | 2005–2016 |
|  | EPA-NCU (EN) | 24.968/121.185 | 144 | 2005–2016 |
|  | Taihu (TH) | 31.421/120.215 | 20 | 2005–2016 |

The AERONET collaboration is a network of ground-based sun photometers, providing globally distributed observations of spectral AOD in the range of 0.34–1.06 μm, with a high temporal resolution (~15 min) (https://aeronet.gsfc.nasa.gov). Datasets with a total of three quality levels: Level 1.0 (unscreened), Level 1.5 (cloud-screened), and Level 2.0 (cloud-screened and quality-assured) are available at the AERONET website. The processing algorithms of AERONET have evolved from Version 1.0 to Version 2.0 and now Version 3.0 (Giles et al., 2017). Thanks to the high accuracy and standardization, the Level 2.0 AERONET AOD data of Version 3.0 with low uncertainty were deemed to be the ground truth to validate the MODIS C6.1 AOD retrievals in our study. Meanwhile, aerosol size distribution ($\alpha_{440-870}$) data were obtained from the Version 3.0 Level 2.0 AERONET AOD product for further analysis, as described in Sections 3.2.1 and 3.2.2. A total of 20 AERONET sites distributed in the north and south of China are listed in Table 1, with diverse surface types and elevations.

*2.1.2 MODIS AOD products in C6.1*

(1) The DT AOD products

In brief, the DT algorithm in C6 takes advantage of the short-wave infrared (SWIR) band (2.13 µm) (which is affected only slightly by aerosol) to identify dark targets, following the exclusion of all the water, cloud, and snow/ice pixels using masks. According to the assumed relationships, the DT algorithm acquires the surface reflectance in four bands (0.47, 0.66, 1.24, and 2.13 µm). The apparent reflectance in the 0.47 µm and 0.66 µm channels is then calculated to retrieve the AOD at 0.55 µm with the help of look-up table (LUT), including coarse and fine particle aerosol models (Levy et al., 2007). All the procedures are applied to individual boxes of 20 × 20 pixels at a 500-m resolution. From the initial version to C6, the major improvements of the DT algorithm have been aimed at improving the surface reflectance relationships among the bands (red, blue, and SWIR) (Kaufman et al., 1997b; Levy et al., 2007), and the improvements in C6.1 are no exception (Levy et al., 2016). Considering the UP and the normalized difference vegetation index based on SWIR bands ($NDVI_{SWIR}$) as the threshold, the DT algorithm in C6.1 revises the surface characterization when the UP is greater than 20%. In our study, the scientific data set (SDS) named "Optical_Depth_Land_And_Ocean" with quality assurance (QA) ≥1 over ocean and QA=3 over land was selected (Levy et al., 2013).(2) The DB AOD products.

Firstly, the DB algorithm was designed to retrieve aerosol properties over arid, semiarid, and urban areas, where the surface reflectance is normally bright in the red and near-infrared bands. However, surface reflectance in such places is much darker in the DB band (0.5 µm) (Hsu et al., 2004). As opposed to the DT algorithm, DB retrieves AOD at a 1-km spatial resolution using the surface reflectance database (0.412, 0.47, and 0.65 µm), and then aggregates pixels to 10 km. After a dozen years, a hybrid method which makes use of the integration of the predetermined surface reflectance database and the normalized difference vegetation index (NDVI) to estimate the surface reflectance has evolved (Hsu et al., 2013). As a result, the AOD coverage produced by the enhanced DB algorithm has been expanded to all land areas except snow and ice. The DB algorithm has now been updated once more. In C6.1 (Hsu, 2017), extensive effort has gone into improving the surface reflectance modeling for rugged and elevated terrain surface types. In our study, we considered the SDS: "Deep_Blue_Aerosol_Optical_Depth_550_Land_Best_Estimate" (QA=2, 3 over land) as the appropriate data for evaluation (Hsu et al., 2013).

**Table 2:** Detailed information about the IGBP land cover classification used in this study.

| Class number | Simplified class name | Specific class name |
|---|---|---|
| 0 | Water | *Water bodies* |
| 1 |  | *Evergreen needleleaf forest* |
| 2 |  | *Evergreen broadleaf forest* |
| 3 |  | *Deciduous needleleaf forest* |
| 4 |  | *Deciduous broadleaf forest* |
| 5 |  | *Mixed forest* |
| 6 | Vegetation | *Closed shrublands* |
| 7 | (high and low elevation) | *Open shrublands* |
| 8 |  | *Woody savannas* |
| 9 |  | *Savannas* |
| 10 |  | *Grasslands* |
| 12 |  | *Croplands* |
| 14 |  | *Cropland/natural vegetation mosaic* |
| 11 | Wetlands | *Permanent wetlands* |
| 13 | Urban | *Urban and built-up* |
| 15 | Snow and ice | *Snow and ice* |
| 16 | Bare soil | *Barren or sparsely vegetated* |

*2.1.3 MODIS land cover type product*

The MODIS land cover type product (MCD12Q1, 500 m) provides data depicting five global land cover classification systems (Friedl et al., 2010), which describe the land cover properties derived from observations spanning a year's input of Terra- and Aqua-MODIS data. The primary land cover scheme employed in our study uses the 17 land cover classes defined by the International Geosphere Biosphere Programme (IGBP). Further details about IGBP land cover types can be found in Table 2.

*2.1.4 The Shuttle Radar Topography Mission (SRTM) product*

The SRTM was a specially modified radar system carried onboard the Endeavour Space Shuttle in the 11-day STS-99 mission between February 11–22, 2000. SRTM successfully obtained radar data over 80% of the Earth's land surface between 60°N and 56°S. There are multiple versions (V1\V2\V3\V4), formats (Hgt\Geotiff\Bil\Arc Grid), and precisions (SRTM1\SRTM3\SRTM30) in the SRTM DEM datasets (http://srtm.csi.cgiar.org). Among the different products, the SRTM1 V3 DEM with a 30-m spatial resolution in China was utilized as the auxiliary terrain data in our study.

*2.2. Methodology*

Since AERONET offers repeated measurements of a point, while satellites provide measurements of a certain region at a single moment. AERONET and MODIS measurements must be matched in space and time. In our study, matchups with the average of the AERONET AOD observations (at least two) at the Terra overpass time (10:30 local time, ±30 mins) and MODIS AOD pixels (at least five) within a radius of 25 km of the AERONET site were adopted (He et al., 2017; Ichoku et al., 2002; Sayer et al., 2014; Tao et al., 2015; Wang et al., 2010). AERONET instruments collect data in multiple wavelengths, many of which are slightly different from the MODIS channel used (550 nm). Therefore, the ground-based AOD at 550 nm was interpolated using the Ångström exponent ($\alpha$), which is defined as shown in Eq.

(1):

$$\alpha = -\frac{\ln(\tau_1/\tau_2)}{\ln(\lambda_1/\lambda_2)} \quad (1)$$

where $\tau_1$, $\tau_2$ represent the AOD at wavelengths $\lambda_1$, $\lambda_2$. The results were validated using the expected error (EE), as shown in Eq. (2) (Sayer et al., 2014); the root-mean-square error (RMSE), as shown in Eq. (3); the mean absolute error (MAE), as shown in Eq. (4); and the relative mean bias (RMB), as shown in Eq. (5):

$$EE = \pm(0.05 + 0.15 \times AOD_{AERONET}) \quad (2)$$

$$RMSE = \sqrt{\frac{1}{n}\sum_{i=1}^{n}(AOD_{(MODIS)i} - AOD_{(AERONET)i})} \quad (3)$$

$$MAE = \frac{1}{n}\sum_{i=1}^{n}|AOD_{(MODIS)i} - AOD_{(AERONET)i}| \quad (4)$$

$$RMB = \overline{AOD_{MODIS}}/\overline{AOD_{AERONET}} \quad (5)$$

**Table 3:** Validation results of the DT retrievals against AERONET sites. The name of specific class can be seen in Table 2. The correlation coefficients (R) were calculated at a significance level of p<0.05. N: number of matched points; EE: expected error of the AOD retrievals; Ele: mean elevation of the matched area.

| Simplified class | Specific class | Site | N | R | RMSE | MAE | RMB | >EE% | :EE% | <EE% | Mean AOD | Ele (m) |
|---|---|---|---|---|---|---|---|---|---|---|---|---|
| Urban | 13 12 | BJ | 860 | 0.958 | 0.18 | 0.14 | 1.12 | 42.09 | 53.49 | 4.42 | 0.58 | 50.7 |
| | | BJC | 289 | 0.945 | 0.19 | 0.14 | 1.09 | 42.56 | 51.56 | 5.88 | 0.54 | 56.1 |
| | | ZSU | 16 | 0.938 | 0.16 | 0.13 | 1.22 | 37.5 | 62.5 | 0 | 0.52 | 6.3 |
| Low elevation vegetation | 12 5 8 13 | HZC | 30 | 0.936 | 0.21 | 0.18 | 1.21 | 46.67 | 50 | 3.33 | 0.67 | 15.7 |
| | 14 2 12 | KP | 8 | 0.908 | 0.06 | 0.05 | 0.98 | 0 | 100 | 0 | 0.49 | 18.7 |
| | 12 13 | PP | 18 | 0.9 | 0.44 | 0.27 | 0.94 | 33.33 | 44.44 | 22.22 | 0.94 | 35.8 |
| | 12 | NU | 44 | 0.938 | 0.26 | 0.23 | 1.28 | 70.45 | 29.55 | 0 | 0.76 | 15.9 |
| | | YP | 9 | 0.969 | 0.26 | 0.19 | 1.07 | 33.33 | 55.56 | 11.11 | 1.06 | 21.1 |
| | | HF | 33 | 0.945 | 0.2 | 0.16 | 1.32 | 51.52 | 48.48 | 0 | 0.5 | 37 |
| | | LN | 11 | 0.96 | 0.17 | 0.1 | 0.91 | 9.09 | 72.73 | 18.18 | 0.43 | 13.1 |
| | | SX | 34 | 0.896 | 0.21 | 0.14 | 1.16 | 29.41 | 67.65 | 2.94 | 0.61 | 24.8 |
| | | XH | 1011 | 0.966 | 0.18 | 0.12 | 1.09 | 29.38 | 66.67 | 3.96 | 0.58 | 16.4 |
| High elevation vegetation | 10 | ABT | 12 | 0.65 | 0.06 | 0.06 | 1.16 | 33.33 | 58.33 | 8.33 | 0.16 | 1339.5 |
| | | YL | 19 | 0.953 | 0.09 | 0.08 | 0.9 | 0 | 78.95 | 21.05 | 0.55 | 1105.8 |
| | | SA | 289 | 0.765 | 0.13 | 0.11 | 1.4 | 55.36 | 44.29 | 0.35 | 0.25 | 1802.6 |
| | 5 12 10 | XL | 365 | 0.945 | 0.11 | 0.07 | 1.12 | 19.45 | 74.79 | 5.75 | 0.23 | 764.8 |
| | 5 2 | LL | 406 | 0.714 | 0.16 | 0.12 | 3.1 | 76.35 | 23.65 | 0 | 0.06 | 2101.1 |
| Water | 0 2 13 11 | HKP | 53 | 0.906 | 0.1 | 0.07 | 1.02 | 11.32 | 83.02 | 5.66 | 0.45 | 102.5 |
| | 0 2 12 13 | EN | 247 | 0.803 | 0.17 | 0.1 | 1 | 23.08 | 65.99 | 10.93 | 0.36 | 121.5 |
| | 0 12 | TH | 227 | 0.879 | 0.34 | 0.3 | 1.45 | 83.7 | 15.86 | 0.44 | 0.62 | 8.7 |
| Total | | | 3981 | 0.946 | 0.19 | 0.13 | 1.17 | 41.9 | 54.03 | 4.07 | - | - |

## 3. Evaluation results and discussion

3.1. Evaluation of the MODIS AOD retrievals in C6.1 for China

3.1.1. Validation of the DT retrievals over regions with multifarious underlying surfaces

Overall, the DT retrievals in C6.1 perform well for a total of 3981 matched points at 20 AERONET sites in China. As can be seen from Table 3, the correlation coefficient (R) is 0.946, while the fraction within the EE can be considered relatively low at only 54.03% and the overestimation is 17% (RMB=1.17). The validation results vary widely with the change of the diverse land cover types and aerosol models among all the sites.

For the urban areas, the relationships of reflectance are usually complicated (Small, 2002). After carrying out the improvement of the urban surface reflectance scheme (Levy et al., 2016), the R value for the three urban sites is around 0.94, indicating that the performance of the DT retrievals in C6.1 for urban areas has made some progress. Apart from the Zhongshan_Univ (ZSU) site, the two sites which are located in Beijing (BJ and BJC) show less overestimation (9% and 12%) compared to the significant overestimation in C6 (Nichol et al., 2016). Detailed information about the distinction between DT C6 and C6.1 AOD retrievals in urban areas can be found in Section 3.2.1.

Aiming at retrieving AOD from low surface reflectance, the DT algorithm embodies its advantage on a large scale in the vegetation areas (Kaufman et al., 1997b; Levy et al., 2007, 2013, 2016). The percentage of R values which exceed 0.93 comes to nearly 60%. Furthermore, the RMSE and MAE reach 0.16 and 0.11, on average, respectively. Although the performance of the overall AOD retrievals deserves to be considered excellent, defects do remain. It can be observed that the DT retrievals generally overestimate the AOD values in vegetation areas. As for the DT algorithm, the error derives from two main sources. One is the surface reflectance, and the other is the aerosol models in the LUT (Bilal et al., 2014, 2016; He et al., 2010). Under normal circumstances, surface reflectance is to blame during low aerosol loadings. However, aerosol models take responsibility during high aerosol loadings (Bilal et al., 2017a). For low elevation sites such as Hangzhou_City (HZC), NUIST (NU), Hefei (HF), and Shouxian (SX), which have above-normal AOD values, the main land cover type is croplands. The obvious phenomena of overestimation emerge due to the improper aerosol models, which are a result of predicting the absorption wrongly (Nichol et al., 2016). The AOD values of the Pku_Pek (PP) and Yufa_Pek (YP) sites are particularly high, and the mean AOD reaches 0.94 and 1.06; nevertheless, the DT retrieval results appear ideal during high aerosol loadings. The performances of the DT retrievals at the three sites of Kaiping (KP), Liangning (LN), and XiangHe (XH) are better than the others, with all the fractions within the EE being more than 66%. High elevation areas tend to be sparsely populated mountains, where aerosol properties are relatively simple. Meanwhile, AOD values at high elevation areas tend to be lower by comparison than low elevation areas. Among the high elevation sites, AOE_Baotou (ABT) is located in Nei Monggol and SACOL (SA) is located in Gansu. From Fig. 2 (a)–(b), it can be seen that the plateau topography of these areas is rugged, and the surface reflectance relationship is not described well. Hence, underestimation of the surface reflectance at these two sites exists, resulting in overestimation of the AOD values (Bilal et al., 2017a; Mei et al., 2014). Meanwhile, the R values are also low. It is worth noting that we discovered greatly overestimated deviation in AOD values at Lulin (LL), as much as twice (RMB=3.1) the ground truth. A variation of the surface reflectance of 0.01 will lead to a change of 0.1 in AOD values (Kaufman et al., 1997a). The bidirectional reflectance distribution function (BRDF) effect at Lulin (LL), which features steep terrain, as shown in Fig. 2 (c), is obvious, likely generating an underestimated deviation in surface reflectance. The effects of surface reflectance are huge during the extremely low aerosol loadings, resulting in great deviation being observed (Bilal et al., 2017a). Apart from Lulin (LL), the AOD retrievals of Yulin (YL) and Xinglong (XL) can be considered favorable, and the fractions within the EE are all above 74%.

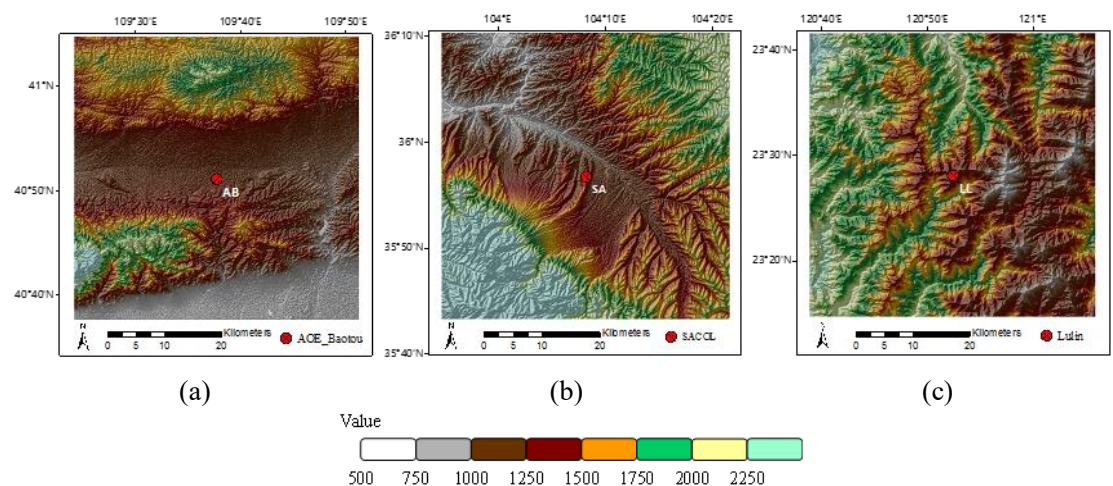

**Fig. 2** DEMs of AERONET sites: (a) AOE_Baotou, (b) SACOL, and (c) Lulin.

As for areas around water, the coastal pixels are removed in the DT land algorithm. Meanwhile, as long as one land pixel exists, then the land algorithm will be adopted (Levy et al., 2013). As the DT algorithm is based on a 10 × 10 km grid, the number of valid pixels in the grids which include coastal pixels will be reduced, likely bringing down the QA (which is based on the number of valid pixels in the window). In addition, C6.1 degraded the QA of the AOD retrievals to zero provided there were more than 50% coastal pixels or 20% water pixels in the 10 × 10 km grid (Mattoo, 2017). Therefore, the valid pixels of the AOD retrievals will be further decreased. Because we employed the "Optical_Depth_Land_And_Ocean" SDS, detailed analysis only on land was difficult to complete. However, it can be found that the Hong_Kong_PolyU (HKP) site shows the best performance, with the fraction within the EE reaching 83.02%. In contrast, overestimation is observed at the Taihu (TH) site, where the RMB is 1.45.

### 3.1.2. Validation of DB against AERONET sites over regions with multifarious underlying surfaces

**Table 4:** Validation results of the DB retrievals against AERONET sites. The name of specific class can be seen in Table 2. The correlation coefficients (R) were calculated at a significance level of p<0.05. N: number of matched points; EE: expected error of the AOD retrievals; Ele: mean elevation of the matched area.

| Simplified class | Specific class | Site | N | R | RMSE | MAE | RMB | >EE% | :EE% | <EE% | Mean AOD | Ele (m) |
|---|---|---|---|---|---|---|---|---|---|---|---|---|
| Urban | 13 12 | BJ | 1646 | 0.944 | 0.17 | 0.11 | 0.98 | 15.49 | 69.44 | 15.07 | 0.47 | 50.7 |
| | | BJC | 482 | 0.948 | 0.14 | 0.09 | 1.01 | 14.52 | 72.41 | 13.07 | 0.39 | 56.1 |
| | | ZSU | 26 | 0.637 | 0.27 | 0.19 | 0.97 | 26.92 | 50 | 23.08 | 0.54 | 6.3 |
| Low elevation vegetation | 12 5 8 13 | HZC | 46 | 0.972 | 0.18 | 0.14 | 0.86 | 4.35 | 52.17 | 43.48 | 0.68 | 15.7 |
| | 14 2 12 | KP | 8 | 0.932 | 0.2 | 0.19 | 0.62 | 0 | 12.5 | 87.5 | 0.49 | 18.7 |
| | 12 13 | PP | 18 | 0.921 | 0.38 | 0.25 | 1.02 | 16.67 | 50 | 33.33 | 0.83 | 35.8 |
| | | NU | 63 | 0.864 | 0.29 | 0.21 | 0.87 | 9.52 | 50.79 | 36.68 | 0.77 | 21.1 |
| | | YP | 8 | 0.965 | 0.32 | 0.22 | 0.97 | 12.5 | 50 | 37.5 | 1.04 | 15.9 |
| | 12 | HF | 51 | 0.942 | 0.13 | 0.1 | 0.94 | 5.88 | 72.55 | 21.57 | 0.56 | 37 |
| | | LN | 19 | 0.916 | 0.16 | 0.12 | 0.92 | 21.05 | 57.89 | 21.05 | 0.37 | 13.1 |
| | | SX | 42 | 0.942 | 0.19 | 0.14 | 0.91 | 4.76 | 66.67 | 28.57 | 0.65 | 24.8 |
| | | XH | 1542 | 0.932 | 0.23 | 0.14 | 1.11 | 34.76 | 57.46 | 7.78 | 0.56 | 16.4 |
| High elevation vegetation | 10 | ABT | 19 | 0.734 | 0.08 | 0.07 | 0.68 | 0 | 68.42 | 31.58 | 0.17 | 1339.5 |
| | | YL | 131 | 0.736 | 0.22 | 0.17 | 0.55 | 1.53 | 26.72 | 71.76 | 0.36 | 1105.8 |
| | | SA | 694 | 0.875 | 0.12 | 0.08 | 0.98 | 15.13 | 70.46 | 14.41 | 0.33 | 1802.6 |
| | 5 12 10 | XL | 611 | 0.865 | 0.15 | 0.06 | 1.07 | 12.27 | 79.87 | 7.86 | 0.2 | 764.8 |
| | 5 2 | LL | 296 | 0.757 | 0.07 | 0.05 | 1.58 | 23.65 | 72.97 | 3.38 | 0.05 | 2101.1 |
| Water | 0 2 13 11 | HKP | 52 | 0.6 | 0.2 | 0.15 | 0.68 | 7.69 | 44.23 | 48.08 | 0.34 | 102.5 |
| | 0 2 12 13 | EN | 69 | 0.835 | 0.13 | 0.09 | 0.99 | 15.94 | 72.46 | 11.59 | 0.39 | 121.5 |
| | 0 12 | TH | 414 | 0.91 | 0.17 | 0.11 | 0.99 | 14.98 | 68.6 | 16.43 | 0.56 | 8.7 |
| | Total | | 6237 | 0.931 | 0.18 | 0.11 | 1.02 | 22.07 | 63.49 | 14.44 | - | - |

Overall, DB retrievals in C6.1 have only a slightly lower R value relative to DT, which reaches 0.931. The other criteria are superior to DT, particularly the RMB, which is close to 1.0. Furthermore, the DB retrievals are able to estimate AOD values with almost no overestimation or underestimation. At the same time, the matched points, which are almost twice as many as DT, are 6237 pairs. Similarly, the validation results vary widely with the different land cover types and aerosol models among the sites.

As for the urban areas of Beijing (BJ) and Beijing-CAMS (BJC), the performance of the DB retrievals is satisfactory. As illustrated in Table 4, the R values for both sites are over 0.94. The fractions within the EE are 69.44% and 72.41%, respectively, suggesting a good estimation. However, the R value only reaches 0.637 and the fraction within the EE is also relatively low at 50% at Zhongshan_Univ (ZSU). This may be caused by the improper selection of aerosol model and the incorrect assumption of surface reflectance (Bilal et al., 2014, 2016; He et al., 2010).

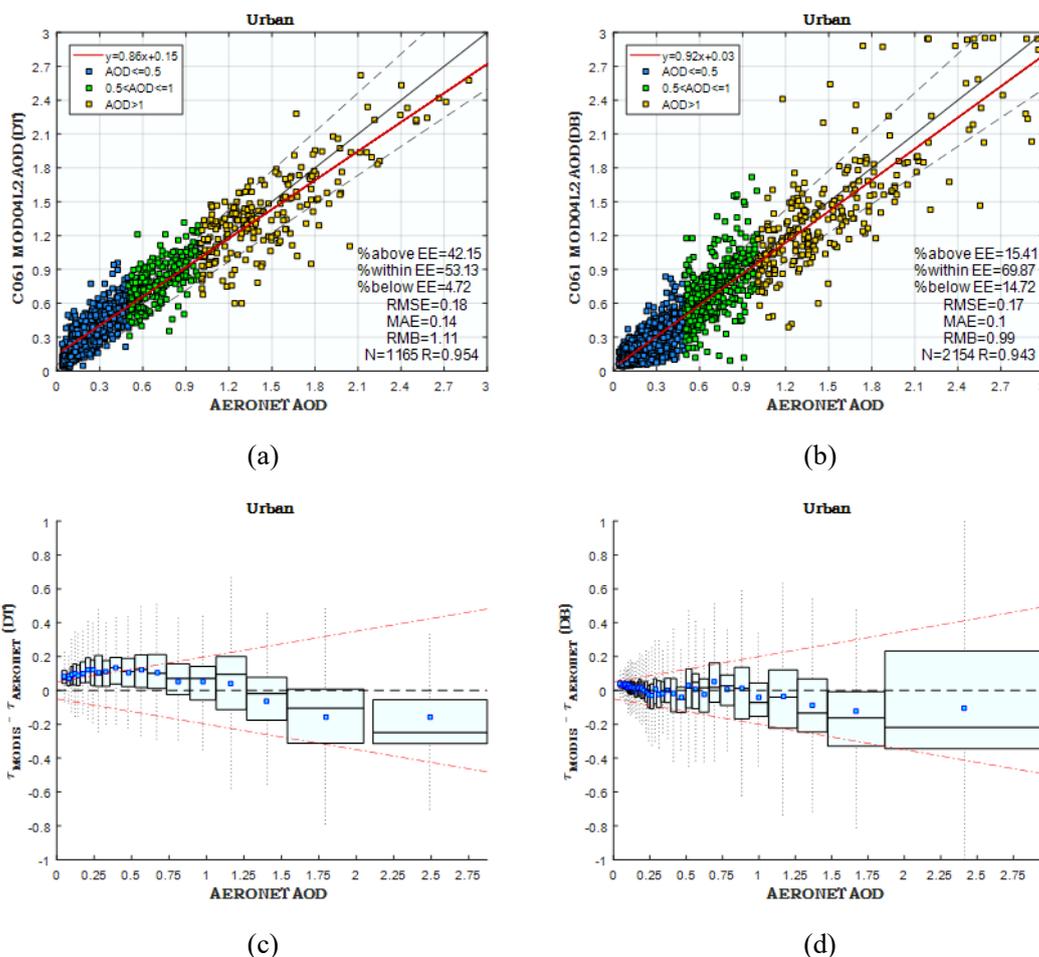

**Fig. 3** Validation results of (a) DT and (b) DB and box diagrams of (c) DT and (d) DB in urban areas. In (a) and (b), the red solid line represents the regression line, the dashed lines are the EE lines, and the black solid line is the 1:1 line. In (c) and (d), the black horizontal dashed line represents zero bias, and the red dotted and dashed lines represent the EE lines. For each box, the middle line, azure dot, and upper and lower hinges represent the AOD bias median, mean, and 25th and 75th percentiles, respectively. The whiskers extend to 1.5 times the interquartile range (IQR).

At vegetation sites with low elevation, the total R value for DB retrievals appears high, but underestimation basically exists. Except the overestimation of 11% (RMB=1.11) at the XiangHe (XH) site and 2% (RMB=1.02) at the Pku_Pek (PP) site, the RMB values for all the other sites are all less than 1.0. It can be inferred that, in these areas, some problems may have emerged in the selection of aerosol model or construction of surface reflectance (Bilal et al., 2014, 2016; He et al., 2010), resulting in regular deviation. We found that the performance at vegetation sites with high elevation is clearly worse than that at sites with low elevation. Meanwhile, this is also reflected in the R values. Apart from the distinct underestimation, significant overestimation is observed at Lulin (LL), with the RMB equal to 1.58. As analyzed in Section 3.1.1, similarly, DB retrievals underestimate the surface reflectance and consequently lead to great overestimation in the retrievals, because of the steep terrain. However, the magnitude of the overestimation is visibly lower than with the DT retrievals.

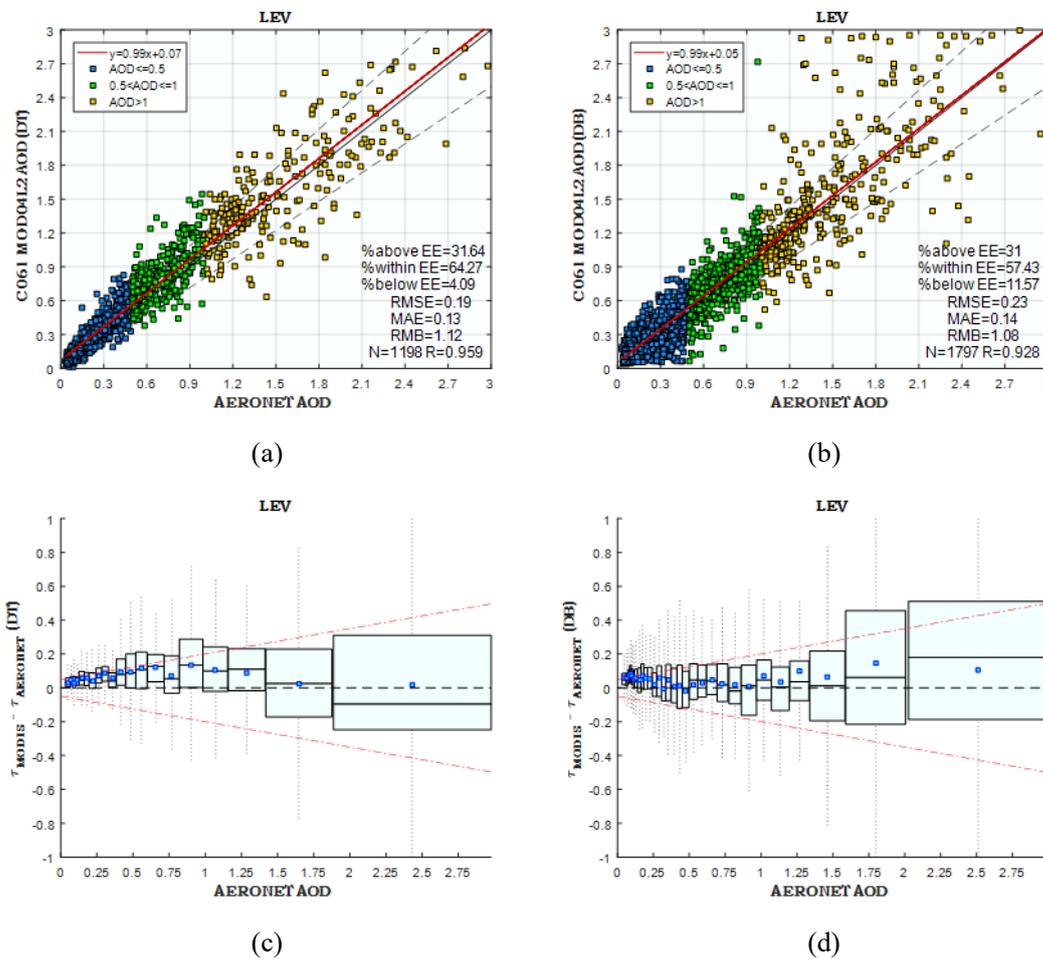

(a) (b)

(c) (d)

**Fig. 4** Validation results of (a) DT and (b) DB and box diagrams of (c) DT and (d) DB in LEV areas. In (a) and (b), the red solid line represents the regression line, the dashed lines are the EE lines, and the black solid line is the 1:1 line. In (c) and (d), the black horizontal dashed line represents zero bias, and the red dotted and dashed lines represent the EE lines. For each box, the middle line, azure dot, and upper and lower hinges represent the AOD bias median, mean, and 25th and 75th percentiles, respectively. The whiskers extend to 1.5 times the interquartile range (IQR).

For areas around water, the quality of DB retrievals is considered well at EPA-NCU (EN) and Taihu (TH) while the R value for EPA-NCU (EN) is lower at 0.835. Surprisingly, it can be seen that the performance of the DB retrievals is extremely poor at Hong_Kong_PolyU (HKP). The R value reaches only 0.6 and obvious underestimation (RMB=0.68) can be observed.

3.1.3. Comparison between DT and DB in accuracy over urban and vegetation areas

After detailed analysis of each site in China, comparisons according to the land cover classification between the DT and DB retrievals were processed. Due to the lack of DB retrievals over water, the land cover types were preliminarily classified as urban and vegetation. Taking the topography effects on AOD retrievals (He et al., 2016; Li et al., 2003; Wang et al., 2010) into account, vegetation was subdivided into high elevation vegetation (HEV) and low elevation vegetation (LEV).

The three sites of Beijing (BJ), Beijing-CAMS (BJC), and Zhongshan_Univ (ZSU) are regarded as urban areas, where the matched points of the DB algorithm (2154) are almost twice as many as DT (1165). The DT algorithm in C6.1 improves the surface reflectance relationship with the UP in urban areas, and the R value for DT (R=0.954) is higher than that of DB (R=0.943), as depicted in Fig. 3 (a)–(b). However, when it comes to RMSE, MAE, and the fraction within the EE, the DB retrievals are superior to DT. At the same time, slight overestimation (RMB=1.11) is observed in the DT retrievals, while the DB retrievals show normal estimation (RMB=0.99). As a result, the overall quality of the DB retrievals is better than DT in urban areas. Next, we present the stability of the AOD retrievals, in the form of box diagrams. Specifically, we sorted the data values of each site in ascending order, and then the data were sampled with an interval of 60. As shown in Fig. 3 (c)–(d), the boxes which show the difference between DB and the AERONET sites remain well within the EE line following the rise of the aerosol loadings. In contrast, the boxes of the difference between DT and the AERONET sites are significantly higher than the EE line when the AOD values are less than 0.5, suggesting that some problems may still exist in the surface reflectance or aerosol models (Bilal et al., 2014, 2016; He et al., 2010) after the improvements in C6.1. With an increase of the AOD values, the mean values of the DB boxes fluctuate. In contrast, an obvious regularity is discovered, in that the mean of the DT boxes is positive on the condition of the AOD values being less than 1.25 followed by negative deviation. In urban areas, the conclusion can be drawn that the DB retrievals are better than DT in stability, with less deviation.

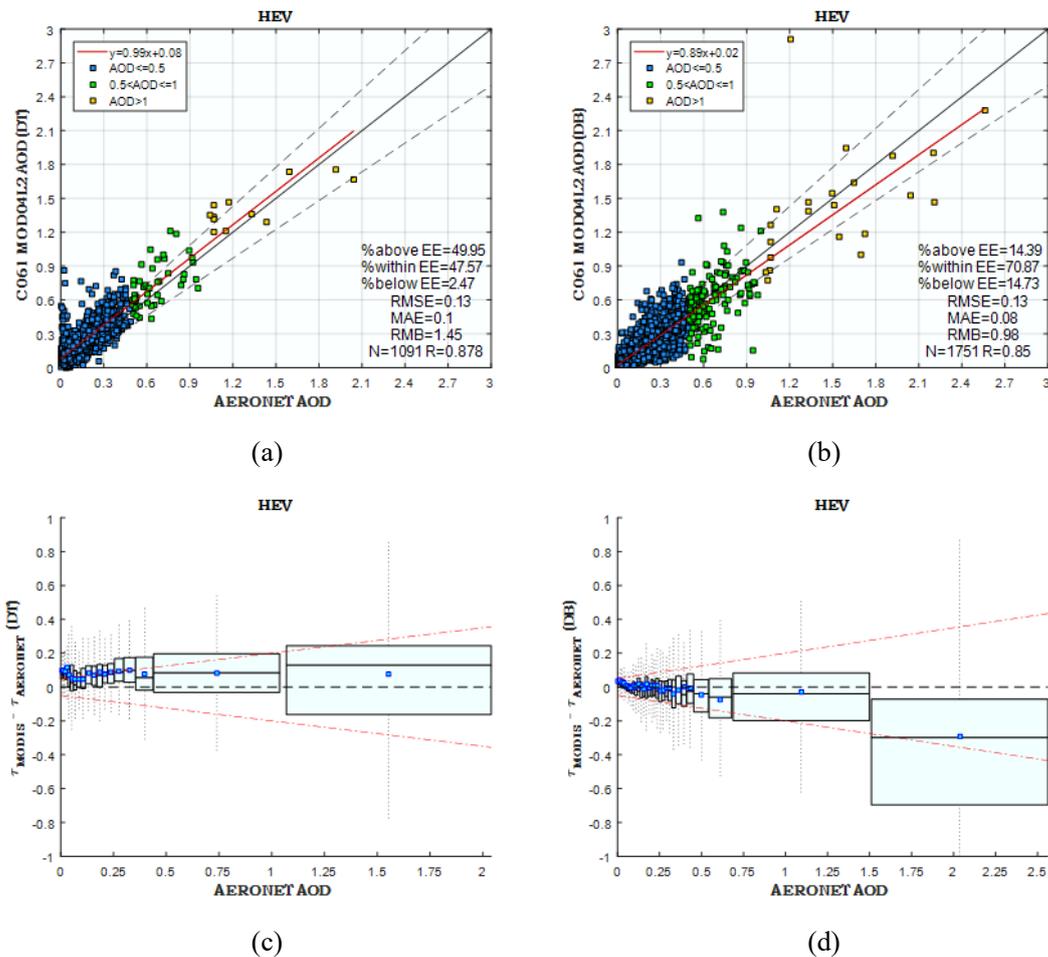

**Fig. 5** Validation results of (a) DT and (b) DB and box diagrams of (c) DT and (d) DB in HEV areas. In (a) and (b), the red solid line represents the regression line, the dashed lines are the EE lines, and the black solid line is the 1:1 line. In (c) and (d), the black horizontal dashed line represents zero bias, and the red dotted and dashed lines represent the EE lines. For each box, the middle line, azure dot, and upper and lower hinges represent the AOD bias median, mean, and 25th and 75th percentiles, respectively. The whiskers extend to 1.5 times the interquartile range (IQR).

The LEV areas consist of a total of nine sites (shown in Table 1) with an elevation of less than 50 m. The main land cover type of each site is croplands. Pixels with surface reflectance in the SWIR band (2.13 μm) greater than 0.1 and less than 0.25 are labeled as dark targets in the DT algorithm (Kaufman et al., 1997b; Levy et al., 2007, 2013, 2016). Only in these areas can DT AOD products be retrieved. In consequence, the DT algorithm is favored in the dense vegetation areas where surface reflectance is up to the standard described above (0.1–0.25). As demonstrated in Fig. 4 (a), except for the slightly higher RMB, the other criteria of the DT retrievals are much better than those of DB in LEV areas. Among the criteria, RMSE and MAE are decreased by 0.04 and 0.01, respectively, and the R value is notably increased by 0.031. Meanwhile, as shown in Fig. 4 (b), when the AOD values are within the range of 1.2–2.4, some overestimated values emerge in the DB retrievals. This may be caused by the improper selection of aerosol model (Bilal et al., 2017a). In conclusion, the performance of the DT retrievals is significantly superior to DB in the LEV areas. As shown in Fig. 4 (c)–(d), what is similar to the urban areas is that the obviously overestimated trend of DT retrievals in the LEV areas still exists as the AOD values are less than 1.5. With regard to the DB retrievals, overestimation is observed on AOD values of less than 0.25. The fluctuation then gradually turns smooth with the rise of the aerosol loading, and overestimation occurs again when the AOD values are greater than 1.0. Overall, the boxes which represent the difference between DB and the AERONET sites stabilize within the EE line when the AOD values are within the range of 0.5–1.5, with a smaller error. Conversely, the boxes fluctuate out of the EE line outside this range.

A total of five sites (shown in Table 1) make up the HEV areas. The elevation ranges from 764.8 m at Xinglong (XL) to 2101.1 m at Lulin (LL). Unlike the LEV areas, the main land cover types in the HEV areas are grassland and forest. It can be observed that the performance of the DT and DB retrievals is inferior to the LEV areas. Although the land cover types are also classified as vegetation, the impact from topography effects on the HEV areas should not be overlooked. As depicted in Fig. 5 (a)–(b), the R value for the DT retrievals is as much as 0.878, and obvious overestimation (RMB=1.45) is apparent. From Sections 3.1.1 and 3.1.2, we speculate that the serious overestimation in DT mainly stems from the contribution of Lulin (LL) during extremely low aerosol loadings. However, for DB retrievals, the RMB, which reaches 0.98, is close to 1, maintaining an excellent level. Meanwhile, the fraction within the EE for DB retrievals (70.87%) is much more than for DT (47.57%), and only the R value is lower at 0.85. Therefore, the performance of the DB retrievals is synthetically better than that of DT in the HEV areas. As shown in Fig. 5 (c)–(d), the boxes showing the difference between DT and the AERONET sites show serious overestimation and are distinctly higher than the EE line on extremely low aerosol loadings. Meanwhile, the overestimation stabilizes when the AOD values increase. In comparison, the box deviation in the DB retrievals is much less, and all the boxes are within the EE line, except the last one. In conclusion, the DB retrievals show little deviation and fluctuation in the LEV areas. However, the deviation of the DT retrievals is always above the zero line.

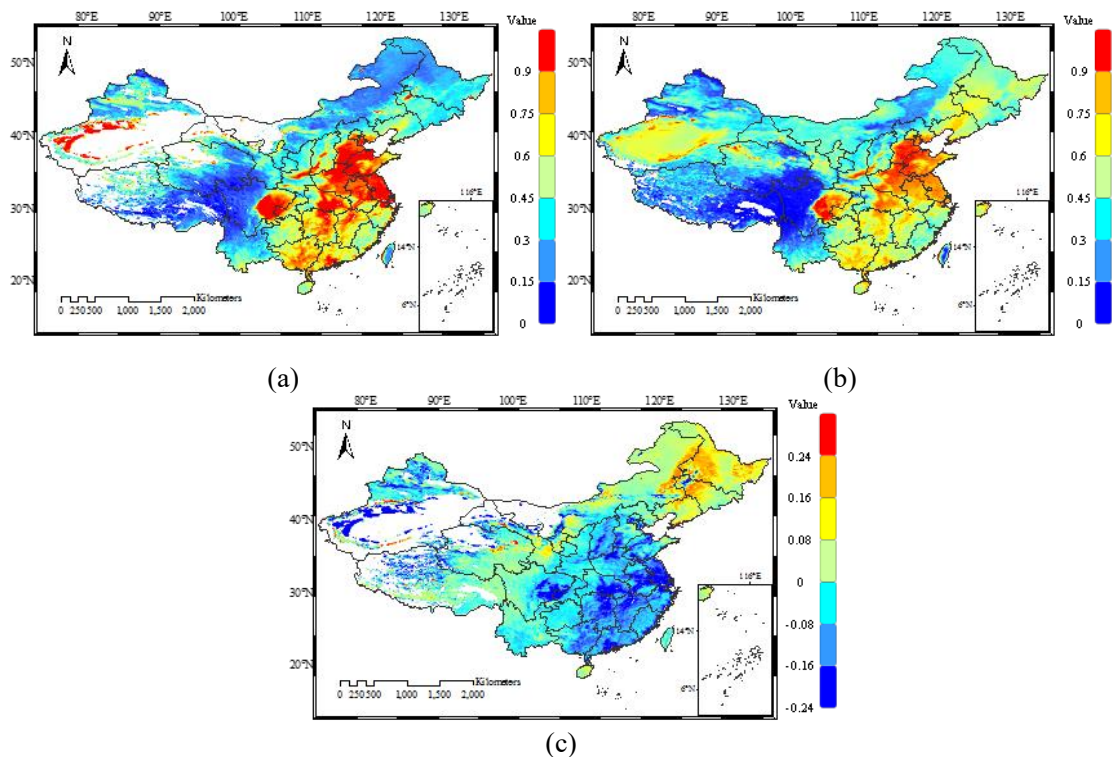

**Fig. 6** 2001–2016 mean AOD retrievals of (a) DT, (b) DB, and (c) DB minus DT over China.

3.1.4. Comparison between DT and DB over regions with multifarious underlying surfaces

Due to the difference in the algorithms, the retrievals of DT and DB are diverse. The difference in retrieved results is a hot issue among scholars, and many people have analyzed the two algorithms in detail (Tao et al., 2015; Xie et al., 2011; Mhawish et al., 2017; Kumar et al., 2018; Sayer et al., 2014; Wang et al., 2010; Bilal et al., 2016; He et al., 2010; Wei et al., 2017). Meanwhile, some researchers have proposed brand-new fusion methods, which are utilized to generate better combined products (Bilal et al., 2017a, 2017b). Furthermore, an official merged dataset was introduced in C6 as early as 2012 (Levy et al., 2013; Sayer et al., 2014). Hence, each algorithm has its pros and cons and cannot replace the other.

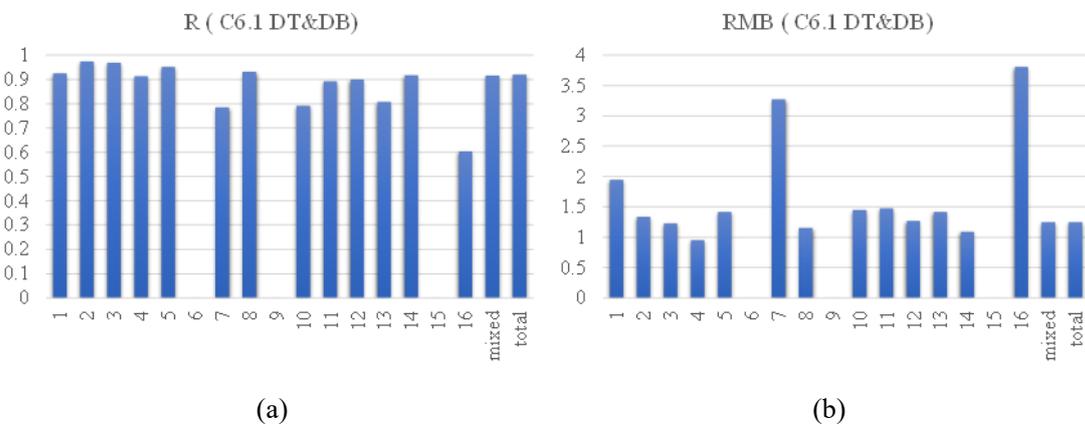

(a)　　　　　　　　　　　　　　(b)

**Fig. 7** (a) Correlation coefficient and (b) RMB of DT and DB classified by land cover types during 2001–2016 over China. The land cover type of each class number in the x-axes can be found in Table 2.

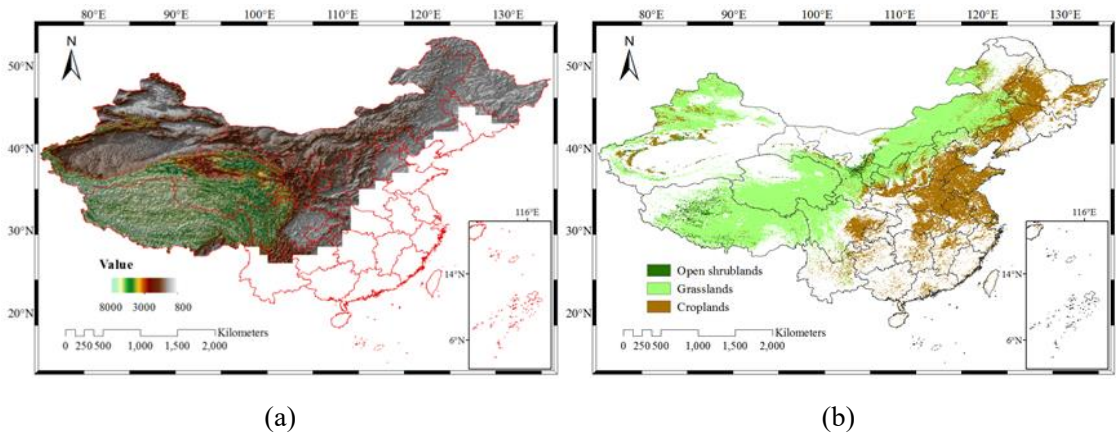

(a)　　　　　　　　　　　　　　(b)

**Fig. 8** (a) DEM and (b) land cover type map (2013) of the target areas.

Firstly, the AOD retrievals from the DT and DB algorithms over China for 2001–2016 were selected for averaging. The mean AOD retrievals of the DT algorithm were then subtracted from the mean retrievals of DB. As shown in Fig. 6 (a)–(b), a long-term of absence of DT retrievals can be observed in Xinjiang, Tibet, and Nei Monggol, where the surface of the local regions is bright. For DB, the AOD retrievals basically cover all the country, except for the areas with snow all year round in Tibet. High aerosol loadings in the DT and DB retrievals occur in east-central China, with a similar spatial structure. As can be seen in Fig. 6 (c), most of the DB AOD values are below the values of the DT retrievals. Meanwhile, the difference during high aerosol loadings is significantly greater than that during low aerosol loadings, on the whole. The diverse retrieval results of both algorithms are mainly caused by the distinction in the selection of aerosol models and the assumption of surface reflectance (Sayer et al., 2014). Sayer et al. (2014) showed that surface reflectance is to blame for the difference in DT and DB during low aerosol loadings. Conversely, aerosol models are responsible during high aerosol loadings. What is worth mentioning is that the majority of the DB AOD values in the northeast are higher than those of the DT retrievals. Since the changes of distribution in surface types of each year are slight at such a large scale, only the land cover type product (2013) was employed for the demonstration. As shown in Fig. 8 (b), the areas where DB is higher than DT basically show the consistent contours of croplands, indicating that croplands may be the main cause of the abnormality. However, crops have also been cultivated in other areas which show normal regularity (DB<DT). Therefore, it can be concluded that the distinction of croplands results in the exception. In China, spring wheat is cultivated in the northeast, while winter wheat is cultivated in the rest of China. With regard to spring wheat, sowing is executed in spring and harvesting is in autumn. As for winter wheat, sowing is completed in late autumn or early winter, and harvesting is in late spring or early summer. The surface reflectance of spring wheat differs from winter wheat in the different periods, while vegetation areas are not subdivided into croplands in the DT algorithm (Kaufman et al., 1997b; Levy et al., 2007, 2013, 2016). Consequently, the distinction of wheat areas may be the most likely reason why the difference between DT and DB in the northeast is different from the other areas.

We then classified the data of the DT and DB retrievals during 2001–2016 into the 16 land cover types and calculated the R value and RMB for each one. Considering that only land AOD products are provided in the DB algorithm, we removed the class of *Water bodies*. Due to the difference in the spatial resolution of the AOD product (10 km) and the land cover type product (500 m), our approach was to take a 20 × 20 pixels grid (10 × 10 km) in the land cover type product and count the number of each class. A grid where more than 90% of the pixels belonged to one class was regarded as the individual class; otherwise, it was regarded as a mixed type. Hence, in addition to the 16 different land cover types, we also added the *Mixedlands* class to fit the mixed type. As shown in Fig. 7 (a)–(b), none of the grids are classified as class 6 (*Open shrublands*), 9 (*Savannas*), and 15 (*Snow and ice*). With regard to the other land cover types, apart from class 7 (*Open shrublands*), 10 (*Grasslands*), 13 (*Urban areas*), and 16 (*Barren or sparsely vegetated*), all the R values between the DT and DB retrievals reach around 0.9, showing strong relevance. In classes 13 (*Urban areas*) and 16 (*Barren or sparsely vegetated*), especially class 16, the DT retrievals are prone to error, which is caused by the high surface reflectance (Kaufman et al., 1997b; Levy et al., 2007, 2013, 2016). However, the DB algorithm is suitable for these areas (Hsu et al., 2013; Hsu, 2017). As a result, the R values between DT and DB tend to be lower than the other places. It can be observed from Fig. 8 (b) that *Open shrublands* mainly distribute in Tibet, and the distribution of *Grasslands* covers not only Tibet, but also Gansu, Qinghai, and Nei Monggol. From the DEM diagram in Fig. 8 (a), most of these areas distribute in plateaus with high elevation. In Sections 3.1.1 and 3.1.2, we noted that some problems emerge in DT and DB retrievals in HEV areas. Consequently, the R values are also reduced to nearly 0.8. As for the *Mixedlands* class, because it accounts for the vast majority of the classes, the total R value basically remains the same at 0.9. Compared with the R value, the RMB turns out to vary greatly only in class 7 (*Open shrublands*) and 16 (*Barren or sparsely vegetated*). As mentioned above, the surface reflectance of class 16 (*Barren or sparsely vegetated*) appears high, and overestimation is found in the DT retrievals, while the DB retrievals remain normal (Hsu et al., 2013; Hsu, 2017; Kaufman et al., 1997b; Levy et al., 2007, 2013, 2016). Class 7 (*Open shrublands*) is mainly found in Tibet, where the terrain is steep and undulating, as can be seen in Fig. 8 (a). Meanwhile, the AOD values are extremely low in Tibet, as demonstrated in Fig. 6 (a)–(b). In Sections 3.1.1 and 3.1.2, we noted that the AOD values of the DT retrievals are significantly overestimated in HEV areas such as Lulin (LL) during extremely low aerosol loadings. In contrast, although the DB retrievals also show overestimation, this is far less than that of DT. Therefore, the RMB for both classes (7 and 16) is high, and even greater than 3. The other land cover types remain at around 1.5, indicating that the AOD values of the DT retrievals are generally higher than those of DB.

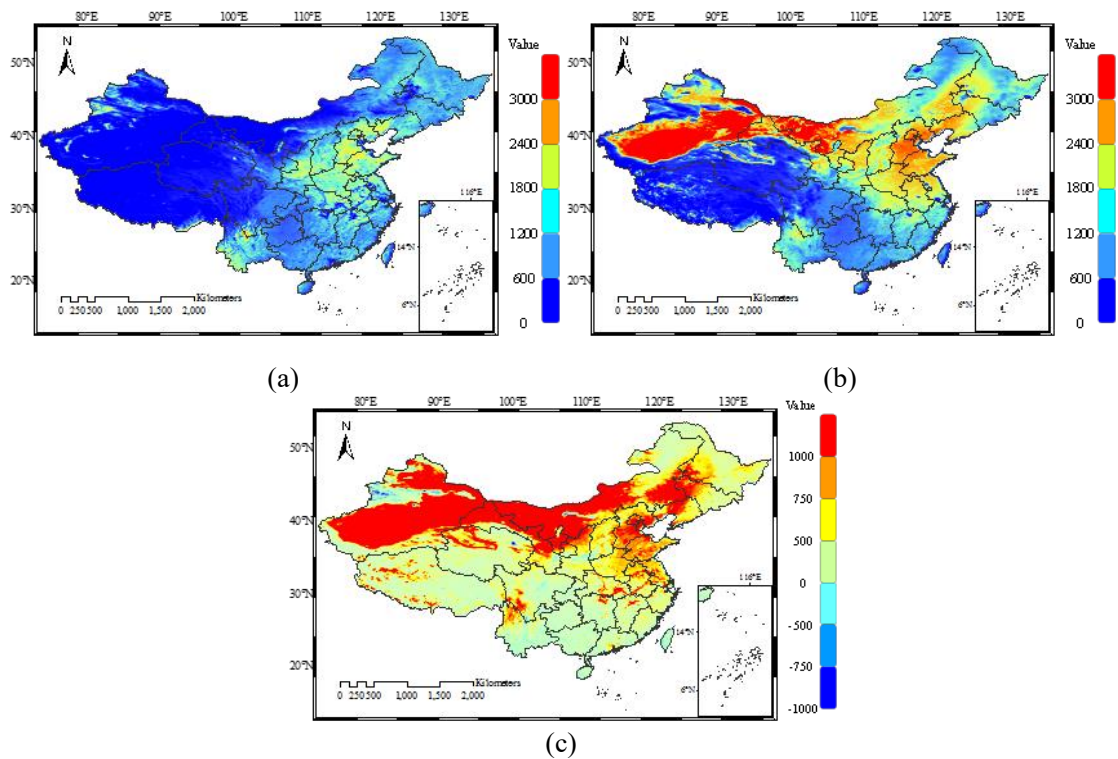

**Fig. 9** 2001–2016 summed AOD coverage of (a) DT, (b) DB, and (c) DB minus DT over China.

In addition to the accuracy of the AOD products, AOD coverage is also a hot spot of concern (Bilal et al., 2017b; He et al., 2017; Tao et al., 2015). Similarly, the AOD coverage data for both the DT and DB algorithms for 2001–2016 were selected to be summed, i.e., the number of valid pixels for each grid. The summed AOD coverage data of the DT algorithm was then subtracted from DT. As depicted in Fig. 9 (c), the total AOD coverage of the DB retrievals exceeds that of DT. From Fig. 9 (a) and (b), the DT retrievals with high coverage mainly distribute in east-central China. In the southwest, the effects of the higher surface reflectance lead to low AOD coverage (Kaufman et al., 1997b; Levy et al., 2007, 2013, 2016). In contrast, the AOD coverage of the DB retrievals tends to be high in Xinjiang, Gansu, and Nei Monggol, where the major land cover type is bare soil. DB retrievals with low coverage center on some areas affected by snow, such as parts of Tibet and Qinghai. Meanwhile, we can also observe that the AOD coverage of the DB retrievals in the northwest of China (bare soil in the majority) is higher than in the south, which we can ascribe to the impact of cloud. As for DB, the main factor resulting in a loss of AOD retrievals is the cloudy areas, where DB fails to retrieve AOD (Hsu et al., 2004; 2013). Areas with bare soil surface are often arid areas when compared to the humid and rainy south, and the probability of cloud is greatly reduced. As a consequence, the ratio of valid AOD pixels also increases. A comparison was then made according to the 16 land cover types, plus the *Mixedlands* class. As shown in Fig. 10, apart from classes 7 (*Open shrublands*), 10 (*Grasslands*), 11 (*Permanent wetlands*), 13 (*Urban areas*), and 16 (*Barren or sparsely vegetated*), there is not much difference in the AOD coverage of the land cover types between the DT and DB retrievals, and the AOD coverage ratio (P) fluctuates at around 50%. As illustrated in Fig. 8 (b) and Fig. 9 (c), the areas where the P value of DB greatly exceeds 50% mainly distribute in Tibet for class 7 (*Open shrublands*) and in the junction of Ningxia, Shaanxi, and Nei Monggol for class 10 (*Grasslands*). With regard to classes 13 (*Urban areas*) and 16 (*Barren or sparsely vegetated*), the AOD coverage of the DB retrievals is higher than that of DT in the places where surface reflectance exceeds the threshold in the DT algorithm (0.25 in 2.13 μm) (Kaufman et al., 1997b; Levy et al., 2007, 2013, 2016). For class 16 (*Barren or sparsely vegetated*) in particular, the P value of DB reaches almost 1.0. Surprisingly, it can be seen that the AOD coverage of the DB retrievals is higher than that of DT in class 11 (*Permanent wetlands*). This may result from the update in the C6.1 DT algorithm, where the QA of the grids with more than 50% coastal pixels or 20% water pixels is degraded to zero (Mattoo, 2017). Only the data of which QA=3 over land and QA≥1 over ocean is adopted in the DT algorithm, generating a great reduction of AOD coverage in class 11 (*Permanent wetlands*). Furthermore, the AOD coverage of DB retrievals accounts for about 55% in total coverage (DT+DB) for the *Mixedlands* class. Overall, the part in which the AOD coverage of DB retrievals exceeds DT accounts for about 12% in total coverage.

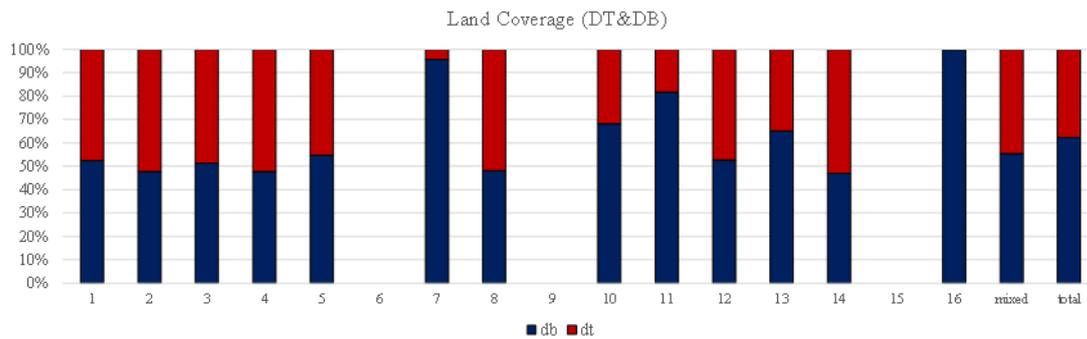

**Fig. 10** AOD coverage ratio of DT and DB classified by land cover type during 2001–2016 over China. The land cover type of each class number in the x-axes can be found in Table 2. The percentage in the y-axes represents the AOD coverage ratio ($\frac{\text{AOD coverage (DT or DB)}}{\text{Total AOD coverage (DT+DB)}}$, P) of each land cover type.

3.2. Comparison between MODIS C6 and C6.1 AOD retrievals in China

3.2.1. Comparison of the DT accuracy over urban areas

C6.1 upgraded the DT algorithm over land using an improved surface relationship. The brand-new surface scheme takes the variation in land cover type into account and is only applied for pixels with a UP of more than 20 % (Levy et al., 2016). MODIS land cover type products are used to identify urban pixels. However, the improved surface ratios for urban areas are generated only using MYD09 products over the continental United States (CONUS) (Levy et al., 2016). Therefore, it is essential to compare DT retrievals over other urban areas between C6 and C6.1. For China, we selected Beijing (BJ) and Beijing-CAMS (BJC), two sites which include a large number of matched points and both of which have a UP of close to 50%. The period for Beijing (BJ) was from 2001 to 2016. With respect to Beijing-CAMS (BJC), the period was from 2013 to 2016. As demonstrated in Fig. 11, the overestimation has been effectively alleviated in C6.1 compared to C6 at Beijing (BJ), at only 12% (RMB=1.12). The fraction within the EE has increased from 37.95% to 53.49%, and the RMSE and MAE have decreased by 0.06 and 0.05. At the same time, the R value has also increased from 0.942 to 0.958, suggesting that the overall performance has been improved significantly. Similar to Beijing (BJ), the RMSE and MAE at Beijing-CAMS (BJC) have decreased, while the R value has gone up. The fraction within the EE has increased by 13.37% and the RMB has decreased by 0.1. Hence, as far as the accuracy is concerned, the applicability of the revised surface reflectance characterization in C6.1 appears extensive. This has resulted in the overestimation caused by the surface reflectance over Beijing being effectively mitigated.

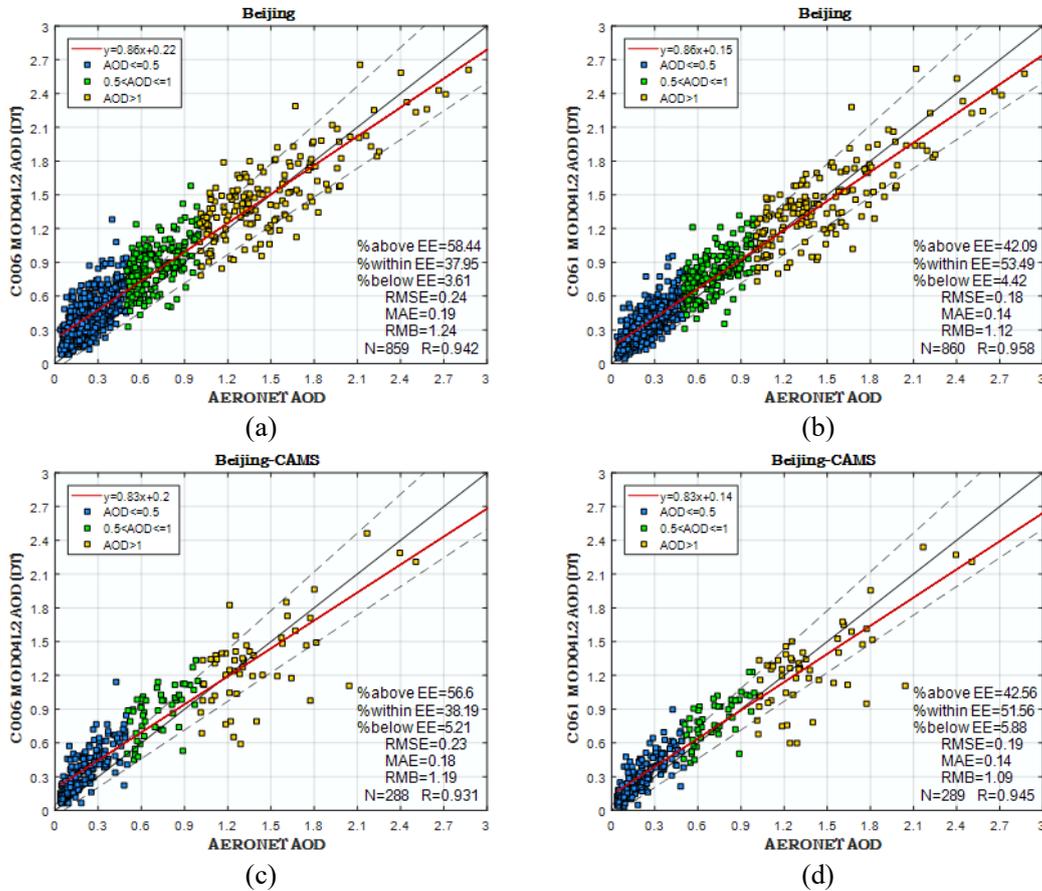

**Fig. 11** Validation results of DT in (a) C6 at Beijing, (b) C6.1 at Beijing, (c) C6 at Beijing-CAMS, and (d) C6.1 at Beijing-CAMS. The red solid line represents the regression line, the dashed lines are the EE lines, and the black solid line is the 1:1 line.

An analysis of the variation trend of DT retrievals between C6 and C6.1 was processed per time series. As can be seen from Fig. 12 (a), the trend of the monthly averaged AOD values shows high correlation between C6 and C6.1 at Beijing (BJ), with consistent rise and fall. Meanwhile, the curve of C6.1 remains almost entirely under that of C6 as time goes by. Similarly, Fig. 12 (b) shows that the monthly deviation between C6.1 and AERONET is clearly less than the monthly deviation between C6 and AERONET, on the whole. What is more, the same conclusion can be drawn from Fig. 12 (c), which represents the daily deviation at Beijing-CAMS (BJC). The daily deviation between C6.1 and AERONET is generally less than the daily deviation between C6 and AERONET. However, at the same time, it occurs to us that a problem emerges. The DT algorithm was modified in C6.1 compared to C6, generating an almost systematic decline in AOD retrievals. With regard to the overestimation, AOD values which are decreased appropriately can increase the accuracy. However, for normal or underestimated results, the action will enlarge the error. Due to the general overestimation in urban areas (Bilal et al., 2016; Nichol et al., 2016), the modification in C6.1 will enhance the overall quality of the DT AOD retrievals. In addition, considering the impact of the aerosol model, we simply divided the aerosol particles at Beijing-CAMS (BJC) into coarse ($\alpha_{440-870}<=0.7$), mixed ($0.7 < \alpha_{440-870}<=1.3$), and fine ($\alpha_{440-870}>1.3$) (Mhawish et al., 2017; Sayer et al., 2014) for further analysis. Compared with the other two types of particle sizes, we can see from Fig. 12 (d) that underestimation is prone to taking place as the aerosol particles are mixed. If the aerosol which creates underestimation in the DT retrievals dominates in the mixed aerosol, the improvement in C6.1 may not be obvious in urban areas. We also ran a sensitivity test for UP, to determine whether the variation of UP would result in modification with diverse levels. Because of the lack of AERONET sites with a variety of UP values, only the comparison between DT AOD retrievals in C6 and C6.1 was made. AOD retrievals from the DT algorithm in C6 and C6.1 for 2001–2016 in China were selected to obtain a series of criteria. From Table 5, we can see that the R value and RMB between C6 and C6.1 go down along with the increase of UP, indicating an adaptive modification. The strength of the modification increases as the UP increases. Meanwhile, it can also be observed that the standard deviation (STD) of C6.1 is less than that of C6, suggesting that the stability of the DT retrievals in C6.1 is better.

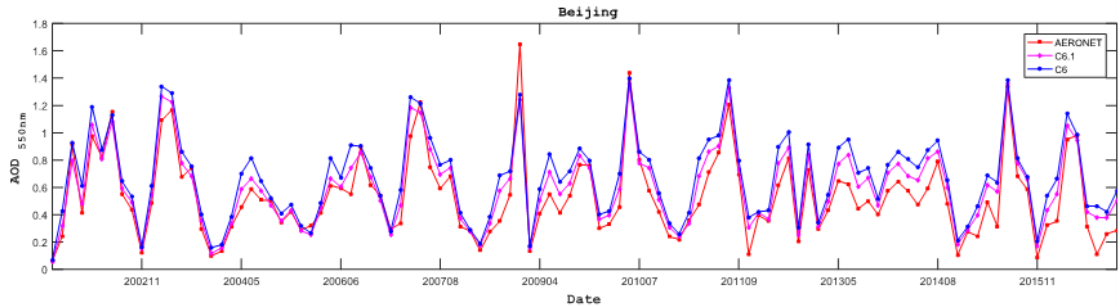

(a)

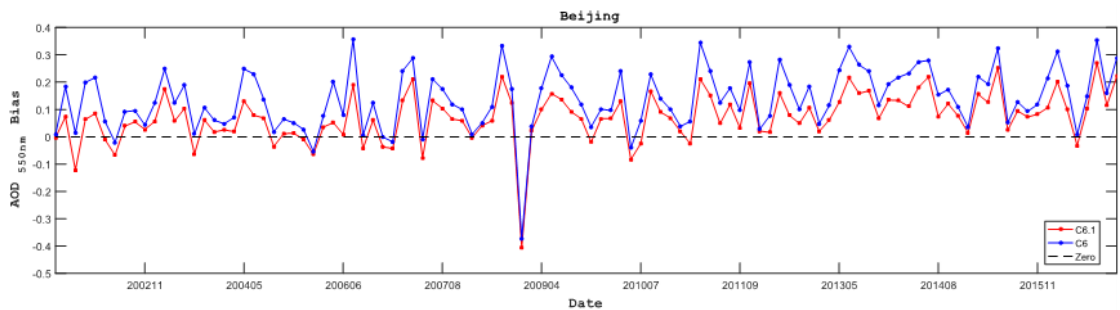

(b)

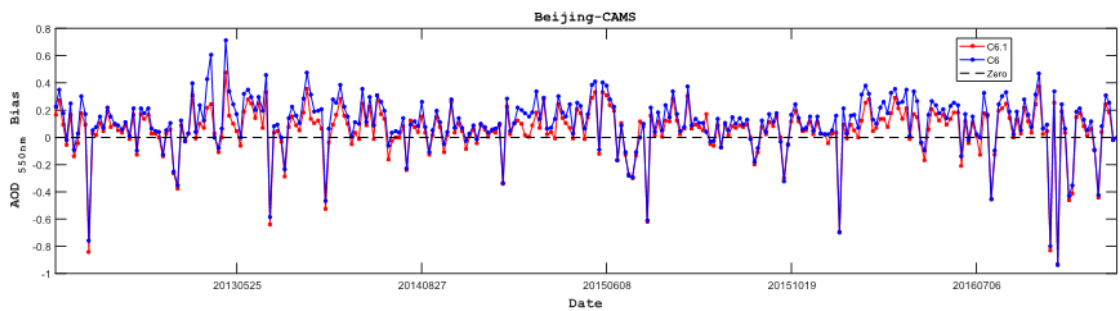

(c)

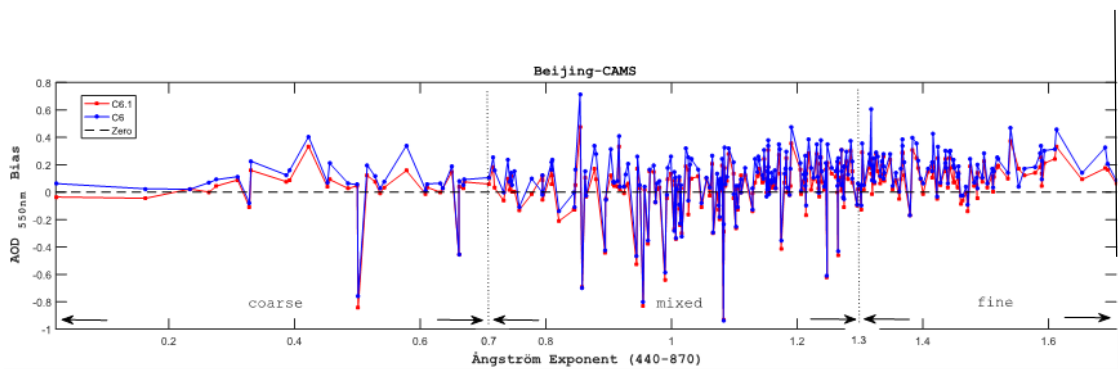

(d)

**Fig. 12** Variation of (a) monthly AOD at Beijing, (b) monthly AOD deviation at Beijing, (c) daily AOD deviation at Beijing-CAMS, and (d) AOD deviation classified by aerosol particles at Beijing-CAMS.

**Table 5:** Sensitivity test results for UP. The light gray represents C6.1 and the dark gray represents C6.

| UP%   | R     | RMB  | MEAN | STD  |
|-------|-------|------|------|------|
| 20-40 | 0.99  | 0.93 | 0.49 | 0.38 |
|       |       |      | 0.52 | 0.38 |
| 40-60 | 0.986 | 0.9  | 0.55 | 0.41 |
|       |       |      | 0.61 | 0.42 |
| 60-80 | 0.977 | 0.87 | 0.57 | 0.42 |
|       |       |      | 0.66 | 0.43 |
| >80   | 0.961 | 0.84 | 0.6  | 0.38 |
|       |       |      | 0.72 | 0.4  |
| Total | 0.98  | 0.89 | 0.55 | 0.4  |
|       |       |      | 0.61 | 0.42 |

3.2.2. Accuracy and AOD coverage of the DB algorithm

Three major improvements of the DB algorithm were introduced in C6.1 (Hsu, 2017) relative to C6. The first is called "heavy smoke detection" (Hsu, 2017). Smoke is thick and homogeneous and can be mistakenly identified as cloud at times, and consequently not retrieved in C6. This gave rise to systematic sampling gaps in the DB AOD retrievals, especially for "weakly absorbing" smoke which appears like cloud in certain spectral bands. The internal smoke detection masks were improved in C6.1 and are capable of addressing some of the over screening events while minimizing true cloud contamination. It is clear that the improvement will raise the number of valid DB pixels. The second aspect is the improved surface modeling in elevated terrain (Hsu, 2017). Terrain with high elevation presents challenges for AOD retrieval from most algorithms and sensors. That is to say, the DB algorithm in C6 always retrieves AOD values that are close to zero under certain conditions in some elevated areas of the world. New surface reflectance models which eliminate the systematic bias were developed in C6.1, resulting in a slight decrease in AOD coverage compared to C6 in these areas. The last improvement is regarded as the artifact reduction in heterogeneous terrain (Hsu, 2017). As it is linked to the underlying terrain, these hotspots are only an issue in certain parts of the world. Therefore, we do not illustrate and discuss the ones in China. In addition, the regional/seasonal aerosol optical models were updated in C6.1.

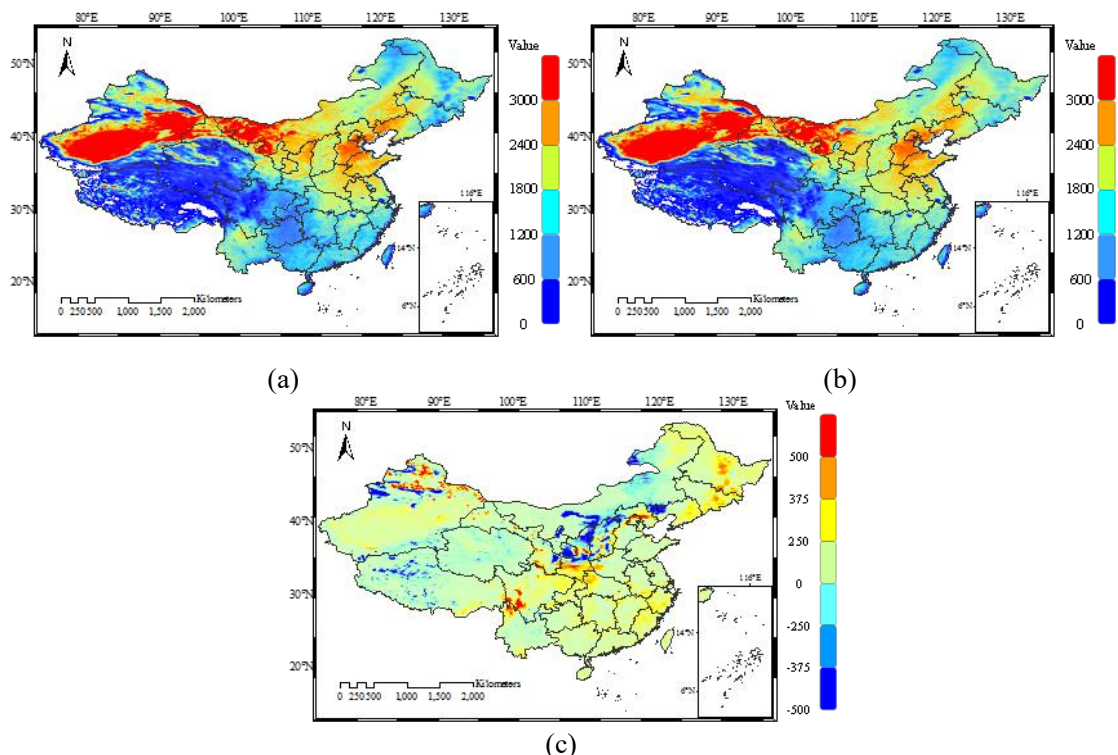

**Fig. 13** 2001–2016 summed AOD coverage of DB in (a) C6, (b) C6.1, and (c) C6.1 minus C6 over China.

Considering the first two improvements, which may have altered the number of valid DB pixels, we compared the DB retrievals of C6 and C6.1 in coverage. To be specific, AOD coverage with the DB algorithm in C6 and C6.1 for 2001–2016 in China were selected to be summed. The summed AOD coverage of the DB algorithm in C6 was then subtracted from C6.1. As shown in Fig. 13 (a)–(b), consistent spatial distribution is observed in the AOD coverage of DB retrievals between C6 and C6.1, with only subtle distinction. Fig. 13 (c) shows that the AOD coverage of C6.1 appears higher than that of C6 in the middle, south, and northeast of China, which is mainly caused by the heavy smoke detection improvement. It can be inferred that this improvement can effectively increase the number of valid DB pixels. Meanwhile, some decreased AOD coverage is found in parts of Tibet, Xinjiang, and the north of China with high elevation. The situation is likely a result of the improved surface modeling in elevated terrain, also indicating that rugged and steep mountains or plateaus distribute in these areas. From Fig. 14 (a)–(b), complex and irregular fluctuation of terrain is discovered in the two places of Gansu where AOD coverage falls off. Nevertheless, we also note some interesting phenomena. As in Nei Monggol in Fig. 14 (c)–(d), the number of valid DB pixels goes down in the relatively flat terrain. This is probably because the area is located within a basin, and the elevation difference between inside and outside tends to be large. Hence, these places are still identified as rugged and steep by the DB algorithm in C6.1. Due to the lack of detailed content in the C6.1 DB algorithm, this does require further study in the future.

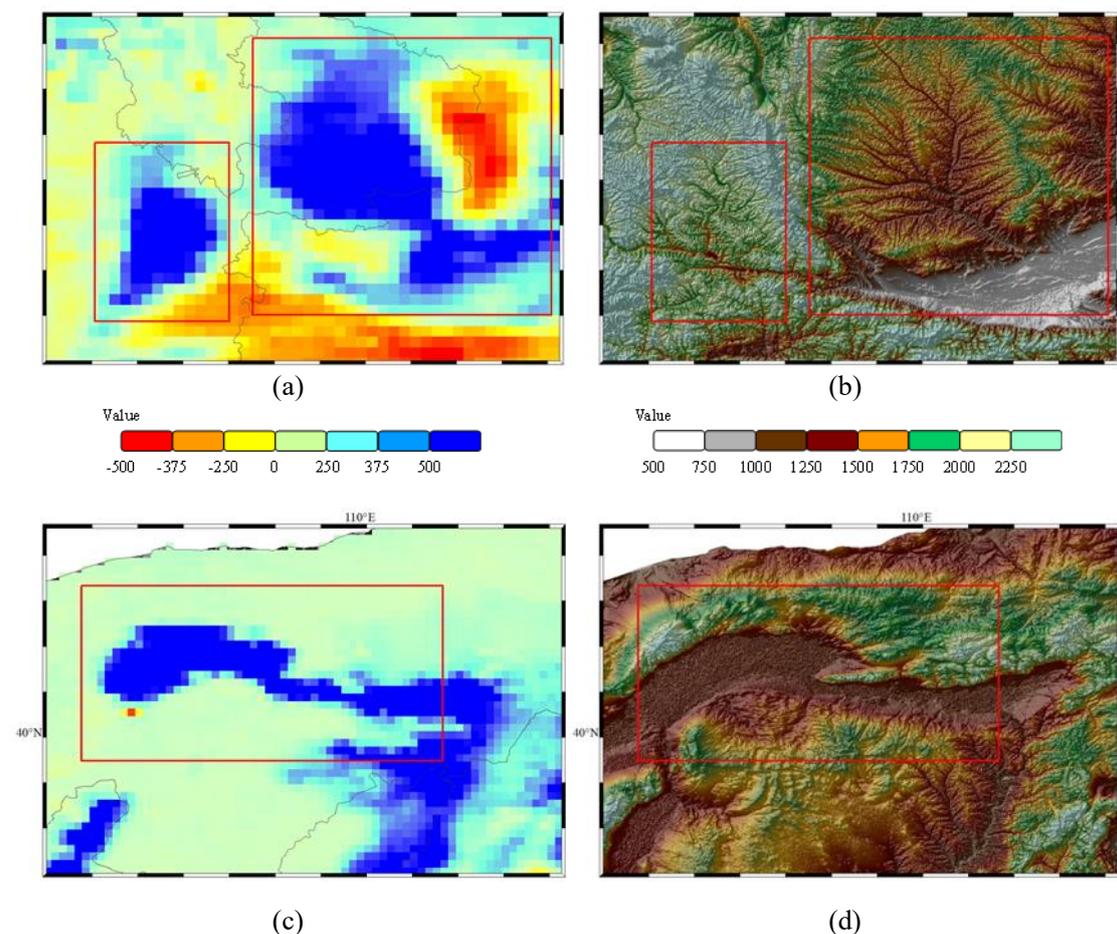

**Fig. 14** 2001–2016 summed AOD coverage of DB and the DEM in the target areas: (a)–(b) Gansu and (c)–(d) Nei Monggol.

In China, no AERONET data are available in the high elevation areas where the DB algorithm has improved the surface modeling in C6.1. Therefore, only the two other improvements (heavy smoke detection and updated regional/seasonal aerosol optical models) could be validated. We integrated the matched points of 20 AERONET sites in China so as to compare the DB retrievals in C6 and C6.1. As depicted in Fig. 15, the quality of the AOD retrievals in C6.1 increases slightly compared to C6. The RMSE and MAE decline by 0.01, and the fraction within the EE goes up from 62.2% to 66.3%. However, at the same time, the RMB and R value show a slight decrease. Overall, the performance of the DB retrievals in C6 and C6.1 is similar. The only significant alteration is that the number of matched points increases from 5883 to 6237, an increment of 6%. In conclusion, the two improvements not only raise the number of valid DB AOD pixels, but also keep the original accuracy. Taking the update of the aerosol models in C6.1 into account, we divided the aerosol particles into coarse ($\alpha_{440-870}<=0.7$), mixed ($0.7<\alpha_{440-870}<=1.3$), and fine ($\alpha_{440-870}>1.3$) (Mhawish et al., 2017; Sayer et al., 2014) for further analysis. From Fig. 16, for both C6.1 and C6, the highest R value emerges for mixed aerosol particles, and the lowest fraction within the EE emerges for coarse aerosol particles. Meanwhile in both C6 and C6.1, when the aerosol particles are coarse, the DB retrievals underestimate the AOD values. Furthermore, we can also observe an overestimation in DB retrievals for fine aerosol particles. Among the three types of aerosol particles, the most remarkable improvement in C6.1 appears for the coarse aerosol particles. After the two improvements are applied, the fraction within the EE is increased from 54.99% to 65.3% and the RMB is increased from 0.88 to 0.95. As for other aerosol particles, the rate of increase is less. It occurs to us that the coarse aerosol particles of the aerosol models in C6 may be inappropriate in China and are suitably corrected in C6.1. Simultaneously, we also discovered that for all types of aerosol particle, the R value in C6.1 always falls relative to C6, to different degrees.

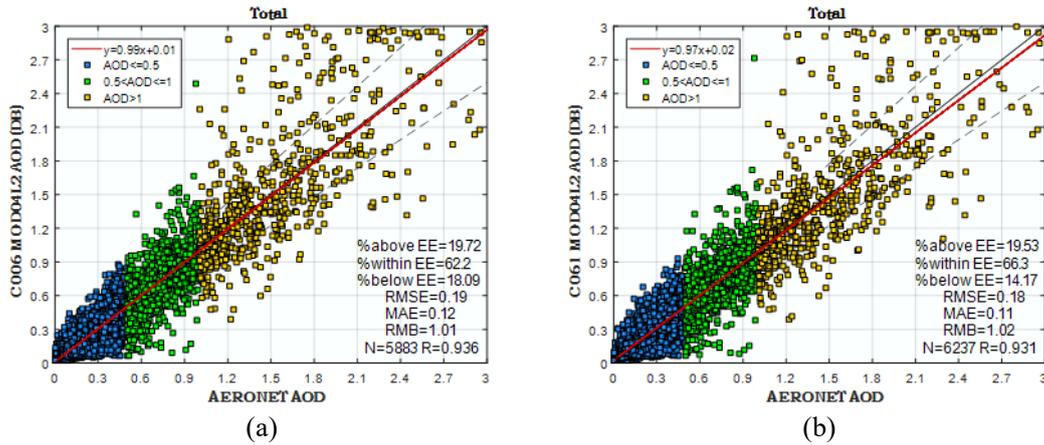

(a)                  (b)

Fig. 15 Validation results of DB in (a) C6 and (b) C6.1 over China. The red solid line represents the regression line, the dashed lines are the EE lines, and the black solid line is the 1:1 line.

## 4. Conclusion

This study was aimed at analyzing the MODIS AOD products. We evaluated the DT and DB retrievals with a 10-km spatial resolution in C6.1 and compared them to C6 over China during 2001–2016. Due to the effects of the diverse surface types and terrains on aerosol properties, a land cover type product (MCD12Q1) and a DEM product (SRTM1 V3.0) were utilized to make the analysis full-scale. The AOD retrievals were validated against Version 3.0 Level 2.0 AERONET AOD products from 20 ground sites in China.

The results show that: 1) The DT retrievals in C6.1 perform well for a total of 3981 matched points at the 20 AERONET sites in China. The R value is 0.946, while the fraction within the EE can be considered relatively low at only 54.03%. The DB retrievals in C6.1 perform slightly better, with a slightly lower R value of 0.931 relative to DT. The other criteria are superior to DT, particularly the RMB, which is close to 1.0. The validation results vary widely with the change of the diverse land cover types and aerosol models among all the sites. 2) Comparing the results over urban and vegetation areas in C6.1, the overall quality of the DB retrievals is better than DT in the urban areas. The performance of DT is significantly superior to DB in the LEV areas. For the HEV areas, DB performs synthetically better than DT. 3) In the spatial distribution aspect in C6.1, most of the DB AOD values are below the values of the DT retrievals. Meanwhile, the difference during high aerosol loadings is significantly greater than that during low aerosol loadings, on the whole. The R values and RMB between DT and DB in C6.1 alter with the different land cover types. 4) For the AOD coverage in C6.1, DT retrievals with high coverage mainly distribute in east-central China. In the southwest, the effects of higher surface reflectance lead to low AOD coverage. In contrast, the AOD coverage of DB tends to be high in Xinjiang, Gansu, and Nei Monggol, where the major land cover type is bare soil. DB retrievals with low coverage center on some areas affected by snow, such as parts of Tibet and Qinghai. Overall, the part in which the AOD coverage of DB exceeds DT accounts for about 12% in total. The AOD coverage ratio (P) also fluctuates with the change of the land cover types. 5) In terms of a comparison of DT between C6.1 and C6, the overestimation in C6 which is caused by surface reflectance is effectively mitigated in C6.1 over urban areas. However, a nearly systematic decline for DT is discovered in C6.1 as well. In addition, the variation of UP in the C6.1 DT algorithm would result in the modification with regard to C6 varying. 6) With respect to DB, between C6 and C6.1, a consistent spatial distribution is observed in AOD coverage, with only subtle distinction. The AOD coverage of DB in C6.1 appears higher than that in C6 in the middle, south, and northeast of China, which mainly results from the heavy smoke detection improvement. The quality of the DB retrievals in C6.1 increases slightly compared to C6, and the most remarkable improvement is observed for the coarse aerosol particles, for which the fraction within the EE is increased from 54.99% to 65.3%.

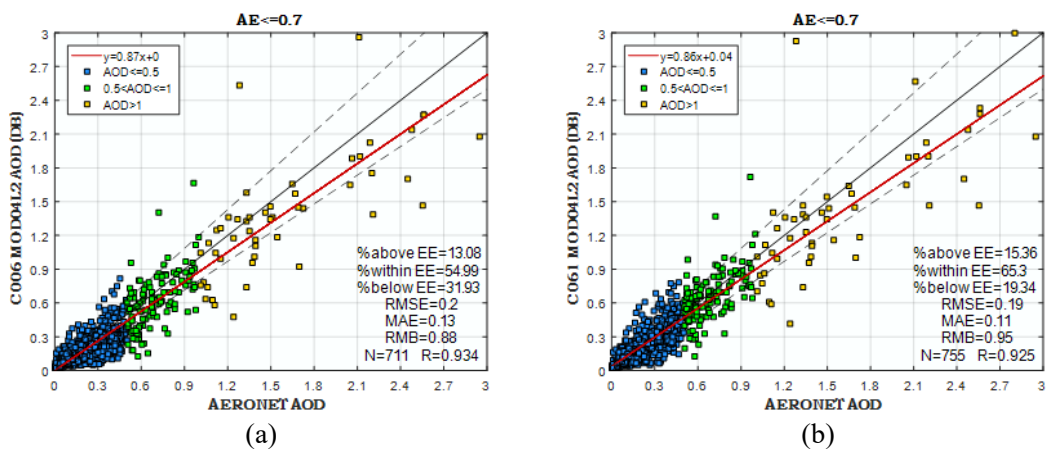

(a)                                   (b)

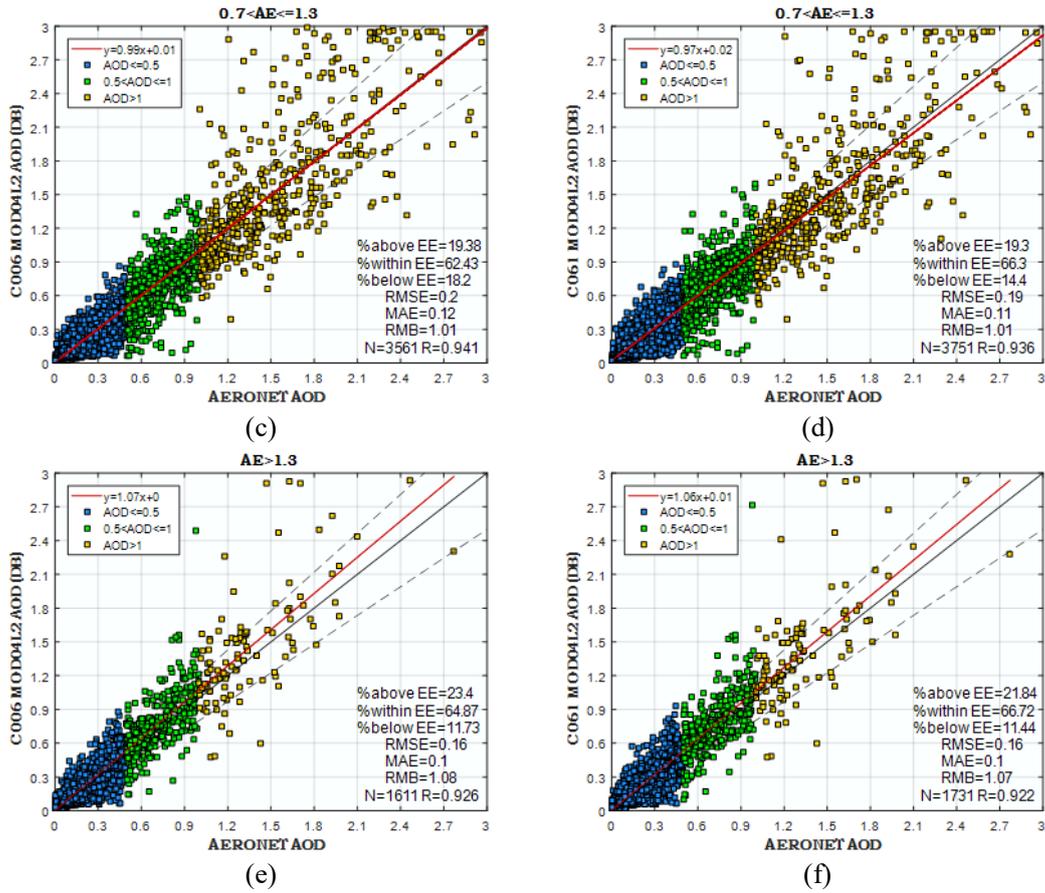

**Fig. 16** Validation results of DB in (a) C6 ($\alpha_{440-870}<=0.7$), (b) C6.1 ($\alpha_{440-870}<=0.7$), (c) C6 ($0.7<\alpha_{440-870}<=1.3$), (d) C6.1 ($0.7<\alpha_{440-870}<=1.3$), (e) C6 ($\alpha_{440-870}>1.3$), and (f) C6.1 ($\alpha_{440-870}>1.3$) over China. The red solid line represents the regression line, the dashed lines are the EE lines, and the black solid line is the 1:1 line.

**Acknowledgments**

This work was supported by the National Key R & D Program of China (No. 2016YFC0200900). The authors would like to express our gratitude to the Atmosphere Archive and Distribution System (LAADS) for providing the MODIS AOD products, and the Principle Investigators for establishing and maintaining the AERONET sites.